\documentclass[preprint,preprintnumbers,amsmath,amssymb,nofootinbib]{revtex4}

\usepackage{graphicx}
\usepackage{color}
\usepackage{changes}
\usepackage{bm}
\usepackage{mathrsfs}
\usepackage{multirow}
\usepackage{array}
\usepackage{float}
\usepackage[colorlinks,citecolor=blue,urlcolor=blue,linkcolor=blue]{hyperref}

\def\bal#1\eal{\begin{align}#1\end{align}}           
\newcommand\bseq{\begin{subequations}}               
\newcommand\eseq{\end{subequations}}
\newcommand\beq{\begin{equation}}                    
\newcommand\eeq{\end{equation}}                       
\newcommand\beqn{\begin{eqnarray}}                   
\newcommand\eeqn{\end{eqnarray}}
\newcommand\nn{\nonumber}                            
\newcommand\fc{\frac}                                
\newcommand\lt{\left}                
\newcommand\rt{\right}
\newcommand\pt{\partial}
\newcommand\tx{\text}
\newcommand\mc{\mathcal}                             

\newcommand{\PreserveBackslash}[1]{\let\temp=\\#1\let\\=\temp}
\newcolumntype{C}[1]{>{\PreserveBackslash\centering}p{#1}}
\newcolumntype{R}[1]{>{\PreserveBackslash\raggedleft}p{#1}}
\newcolumntype{L}[1]{>{\PreserveBackslash\raggedright}p{#1}}

\begin{document}

\allowdisplaybreaks

\title{Exact black hole solutions in Palatini Kalb-Ramond gravity}
\author{
Lin-Yi She$^{a}$,
Rui-Jie Niu$^{a}$,
Ke Yang$^{a}$\footnote{keyang@swu.edu.cn, corresponding author}
}

\affiliation{
$^{a}$School of Physical Science and Technology, Southwest University, Chongqing 400715, China}

\begin{abstract}

We investigate static and spherically symmetric black holes in Palatini Kalb-Ramond gravity, where a nonzero vacuum expectation value of the Kalb-Ramond field spontaneously breaks local Lorentz symmetry. Treating the metric and the torsion-free affine connection as independent variables, we derive the modified field equations and obtain an exact family of solutions encompassing positive, vanishing, and negative values of the cosmological constant $\Lambda$. Unlike the corresponding purely metric formulation, the Palatini theory consistently allows both Lorentz-violating parameters $\ell_1$ and $\ell_2$ to be nonzero. Using the Iyer-Wald formalism, we determine the conserved mass of the black hole solutions. For the $\Lambda\leq0$ sector, we further verify the first law and the Smarr relation, identifying the conserved mass as the thermodynamic enthalpy. We then confront the $\Lambda=0$ solution with Mercury's perihelion precession, solar light deflection, and the Shapiro time delay. Once the central mass parameter is calibrated by planetary dynamics, these observables constrain the Lorentz-violating parameters only through $q\equiv\ell_2/(1-\ell_1)$. Among these tests, Mercury's perihelion precession yields the most stringent constraint, $-7.4\times 10^{-12}\leq q \leq 3.7\times10^{-11}$.

\end{abstract}


\maketitle




\section{Introduction}

The pursuit of a unified theory reconciling General Relativity (GR) and the Standard Model (SM) of particle physics remains a central challenge in theoretical physics. Since GR and the SM are fundamentally incompatible in both concept and mathematics, searching for low-energy signatures that deviate subtly from their predictions provides a highly productive paradigm for probing quantum gravity. One such potential signature is Lorentz symmetry breaking (LSB). Although LSB may be introduced either explicitly or spontaneously, explicit breaking is generally incompatible with standard Riemannian geometry. To ensure geometrical and physical consistency, spontaneous LSB represents the preferred mechanism in gravitational scenarios. In this framework, the Lorentz-violating background emerges as the vacuum expectation value (VEV) of a dynamical tensor field driven by a suitable potential, thereby naturally preserving the covariant conservation of the energy-momentum tensor on-shell. Indeed, spontaneous LSB can arise naturally in string theory \cite{kostelecky1989spontaneous,kostelecky1996expectation}.

To systematically study these symmetry-breaking effects, the Standard Model Extension (SME) was developed as a comprehensive effective field theory framework \cite{colladay1998lorentzviolating,kostelecky2004gravity}. Although its nongravitational sectors have been extensively investigated and constrained (see, e.g., Refs.~\cite{Kostelecky2011,Tasson2014}), the exploration of its gravitational sector has recently attracted growing attention. Among the concrete models relevant to this sector, the bumblebee \cite{bailey2006signals,bluhm2008constraints} and Kalb-Ramond (KR) \cite{altschul2010lorentz} models provide two widely studied realizations of spontaneous LSB. In these formulations, the Lorentz-violating background is generated by the vacuum condensation of a dynamical vector field and an antisymmetric tensor field, respectively.

In bumblebee gravity, a wide variety of black hole solutions have been constructed, including static and spherically symmetric configurations \cite{casana2018exact,maluf2021black,xu2023static,Liu2025d}, charged and slowly rotating charged solutions \cite{Liu2025e}, and rotating black holes \cite{ding2021slowly,ding2023rotating}. More elaborate configurations include black holes with topological defects \cite{gullu2022schwarzschildlike}, dyonic charges \cite{li2025dyonic}, and Taub-NUT-like geometries \cite{li2025dyonic,chen2025taubnutlike}. Their physical properties and observational signatures have also been extensively investigated, including extended thermodynamics and phase transitions \cite{mai2023extendeda}, covariant thermodynamics and black hole evaporation \cite{An2024}, and GUP-modified Hawking radiation \cite{kanzi2019gup}. Astrophysical applications encompass black hole shadows \cite{wang2021shadow}, gravitational lensing \cite{Kumar2026}, accretion processes \cite{cai2023accretiona,yang2019effectsa}, quasinormal modes \cite{oliveira2021quasinormal,chen2023quasinormal,sekhmani2025quasinormal,liu2023qnms}, and quasi-periodic oscillations \cite{Wang2022}. Bumblebee field induced Lorentz violation has further been explored in traversable wormholes \cite{ovgun2019exacta}, gravitational-wave phenomenology \cite{liang2022polarizationsa,amarilo2024gravitational}, and cosmological settings \cite{maluf2021bumblebee,Lai2026}.

Analogously, KR gravity provides another elegant realization of spontaneous LSB through a rank-2 antisymmetric tensor field $B_{\mu\nu}$ \cite{altschul2010lorentz}. Within this framework, several exact black hole solutions have been constructed, including static and neutral configurations \cite{lessa2020modifieda,yang2023static,Liu2024a,yu2025revisiting}, electrically charged black holes \cite{duan2023electrically,yang2026charged}, and slowly rotating solutions \cite{liu2025shadowa}. Black holes incorporating a global monopole have also been obtained \cite{Belchior2025}. Their physical properties have been investigated through black hole thermodynamics and phase structure \cite{yang2023static,duan2023electrically,Liu2024a,yu2025revisiting,yang2026charged}, shadows \cite{kumar2020gravitational,duan2023electrically,liu2025shadowa,Belchior2025}, light deflection \cite{atamurotov2022particle,kumar2020gravitational,Belchior2025}, and particle dynamics and circular orbits \cite{atamurotov2022particle,duan2023electrically}. KR field induced Lorentz violation has further been explored in traversable wormholes \cite{lessa2021traversable,maluf2022exact} and cosmological settings \cite{maluf2022bianchi}.

While purely metric formulations have proven highly successful, the metric-affine formalism, which is defined by independent metric and affine structures, offers a compelling alternative \cite{Hehl1995}. Physically, the metric-affine framework provides a more general geometric arena by naturally accommodating non-metricity and torsion. These non-Riemannian degrees of freedom are widely expected to manifest in high-energy regimes and play a crucial role in candidate theories of quantum gravity \cite{Pagani2015}. Within this broader framework, a particularly prominent approach is the Palatini formulation, wherein the connection is conventionally assumed to be symmetric (torsion-free), and additionally, matter fields are assumed to couple directly to the metric alone. For GR minimally coupled to standard fields, the metric and Palatini formulations yield equivalent field equations. However, this equivalence breaks down in the presence of nonminimal couplings or more general geometric extensions, leading to physically distinct theories. In the presence of nonminimally coupled LSB fields, the two formalisms lead to distinct sets of field equations, rendering them physically inequivalent theories. 

The bumblebee model has also been studied in the Palatini formulation, within which exact static and stationary axisymmetric black hole solutions have been obtained \cite{filho2023vacuum,filho2024exact}. The phenomenology of these non-Riemannian geometries includes modifications to scattering cross sections \cite{heidari2024scattering}, black hole shadows and gravitational lensing \cite{jha2024constrain,gao2024gravitational}, particle creation \cite{filho2025how}, and neutrino flavor oscillations \cite{shi2025neutrinoa}. More recently, their implications have been extended to gravitational-wave propagation \cite{filho2026gravitational} and quantum vacuum effects, including the Casimir effect \cite{cabral2026casimir}.

In this work, we aim to construct exact static spherically symmetric spacetime solutions within the Palatini formulation of KR gravity. Notably, we highlight a fundamental distinction between the Palatini and purely metric formulations of this theory. As pointed out by Ref.~\cite{yu2025revisiting}, within the purely metric formulation, the Lorentz-violating parameters $\ell_1$ and $\ell_2$ (originating from the nonminimal couplings to the Ricci scalar and Ricci tensor, respectively) cannot both be nonzero for static, spherically symmetric solutions. In contrast, we show that this restriction is absent in the Palatini formulation, thereby allowing a broader class of solutions. Because the conventional Arnowitt-Deser-Misner (ADM) and Komar prescriptions are not directly applicable to these nonminimally coupled Lorentz-violating geometries \cite{An2024,li2025dyonic,yu2025revisiting}, we employ the Iyer-Wald formalism to determine the conserved mass of the resulting black holes. Finally, we use Solar System observations to constrain the Lorentz-violating parameters.

The paper is organized as follows. Section~\ref{Framework} introduces the theoretical framework and field equations. Sections~\ref{Solution} and \ref{Wald_Mass} present the exact solutions and determine their conserved mass, respectively. Section~\ref{Thermodynamics} examines the extended thermodynamics, and Sec.~\ref{Tests} derives the Solar System constraints. Finally, Sect.~\ref{Conclusions} summarizes our results.


\section{KR gravity with palatini formalism}\label{Framework}

We begin with the action for gravity nonminimally coupled to a self-interacting KR field \cite{altschul2010lorentz,lessa2020modifieda}
\bal
S  &= \fc{1}{2\kappa}\int{}d^4x\sqrt{-g}\bigg[R \lt(\Gamma\rt) -2\Lambda  + \xi_1 B^{\mu\nu} B_{\mu\nu} R \lt(\Gamma\rt) + \xi_2 B^{\rho\mu} B^{\nu}{}_{\mu} R_{\rho\nu} \lt(\Gamma\rt) \nn \\
& - \fc{1}{6} H^{\mu\nu\rho} H_{\mu\nu\rho} - V\lt(X\rt) \bigg] + \int{d^4x\sqrt{-g}\mathcal{L}_M},
\label{Main_Action}
\eal
where $\kappa \equiv 8\pi G$ with the Newtonian gravitational constant $G=1$ in geometric units, $\Lambda$ is the cosmological constant, $X \equiv B^{\mu\nu}B_{\mu\nu}\pm b^2$, and $\mathcal L_M$ is the matter Lagrangian. The parameters $\xi_1$ and $\xi_2$ are the coupling constants governing the nonminimal interactions between gravity and the KR field. The field strength of the KR field is defined as $H_{\mu\nu\rho} \equiv \partial_{[\mu} B_{\nu\rho]}$. Within the Palatini framework, the spacetime metric $g_{\mu\nu}$ and the symmetric connection $\Gamma^{\lambda}{}_{\mu\nu}$ are treated as independent field variables, with the matter fields coupling exclusively to the metric. Accordingly, all tensor indices are raised and lowered using the spacetime metric $g_{\mu\nu}$, which is also used to contract the Ricci tensor to form the Ricci scalar $R(\Gamma) \equiv g^{\mu\nu}R_{\mu\nu}(\Gamma)$.

To trigger spontaneous LSB, the potential $V(X)$ has to generate a nonzero VEV for the KR field. We denote the corresponding VEV by $\langle B_{\mu\nu}\rangle\equiv b_{\mu\nu}$, which satisfies the constant-norm condition, $b^{\mu\nu}b_{\mu\nu}=g^{\mu\alpha}g^{\nu\beta}b_{\alpha\beta}b_{\mu\nu}=\mp b^2$, as is customary in bumblebee and KR gravity models \cite{casana2018exact,lessa2020modifieda}. The nonzero tensor background $b_{\mu\nu}$ selects preferred local directions and thereby spontaneously breaks local Lorentz symmetry. For black hole solutions with a nonvanishing cosmological constant, however, the simultaneous conditions $V=0$ and $V'\equiv\partial V/\partial X =0$ are generally incompatible with the field equations. Following Ref.~\cite{maluf2021black}, we consider the Lagrange-multiplier potential $V=\lambda X$. Variation with respect to $\lambda$ imposes $X=0$, and hence $B^{\mu\nu}B_{\mu\nu}=\mp b^2$. On shell, $V=0$ and $V'=\lambda$, with the value of $\lambda$ determined by the remaining field equations \cite{bluhm2008spontaneous}.

Relative to a unit timelike vector $v^\mu$, the KR field can be decomposed as $B_{\mu\nu}=\tilde E_{[\mu}v_{\nu]}+\epsilon_{\mu\nu\alpha\beta}v^\alpha\tilde B^\beta$, where $v^\mu v_\mu=-1$ and $\epsilon_{\mu\nu\alpha\beta}$ is the Levi-Civita tensor of $g_{\mu\nu}$. The spacelike fields $\tilde E_\mu$ and $\tilde B_\mu$, satisfying $\tilde E^\mu v_\mu=\tilde B^\mu v_\mu=0$, represent the pseudo-electric and pseudo-magnetic components, respectively. In this work, we restrict our attention to a purely pseudo-electric VEV configuration, where $b_{10} = -b_{01} \equiv \tilde{E}(r)$ are the only nonvanishing components. This configuration identically satisfies the constant-norm condition $b^{\mu\nu}b_{\mu\nu} = -b^2$ and ensures the vanishing of the field strength, $H_{\lambda\mu\nu} \equiv 0$.

\subsection{Connection Field Equation}

Variation of the action \eqref{Main_Action} with respect to the independent connection yields
\beq
\nabla^{(\Gamma)}_{\alpha}
\big[\sqrt{-g}\lt[\lt(1+u\rt)g^{\mu\nu}+s^{\mu\nu}\rt]\big]=0,
\label{Connection_Eq}
\eeq
where $\nabla^{(\Gamma)}_\alpha$ denotes the covariant derivative associated with $\Gamma^{\lambda}{}_{\mu\nu}$, and we have defined $u\equiv\xi_1B_{\mu\nu}B^{\mu\nu}$ and $s^{\mu\nu}\equiv\xi_2B^{\mu\rho}B^\nu{}_{\rho}$ for compactness. To cast this equation into a metric-compatible form, we introduce an auxiliary metric $h_{\mu\nu}$ through its densitized inverse,
\beq
\sqrt{-h}\,h^{\mu\nu}
\equiv
\sqrt{-g}\left[(1+u)g^{\mu\nu}+s^{\mu\nu}\right],
\label{Auxiliary_metric_definition}
\eeq
where $h\equiv\det(h_{\mu\nu})$. Consequently, Eq.~\eqref{Connection_Eq}  becomes
\beq
\nabla^{(\Gamma)}_{\alpha}
\left(\sqrt{-h}\,h^{\mu\nu}\right)=0,
\eeq
which is the metric-compatibility condition for $h_{\mu\nu}$. Since the independent connection is torsion-free, it is therefore the Levi-Civita connection of $h_{\mu\nu}$,
\beq
\Gamma^{\lambda}{}_{\mu\nu} = \fc{1}{2} h^{\lambda\rho} \lt( \partial_{\mu} h_{\nu\rho} + \partial_{\nu} h_{\rho\mu} - \partial_{\rho} h_{\mu\nu} \rt).
\label{Gamma_hdd}
\eeq

Solving the defining relation \eqref{Auxiliary_metric_definition} for the inverse auxiliary metric gives
\beq
h^{\mu\nu} = \fc{1}{\sqrt{\det \lt [(1+u)\hat{I} + \hat{s} \rt] }} g^{\mu\alpha} \lt[ (1+u) \delta^{\nu}{}_{\alpha} + s^{\nu}{}_{\alpha} \rt],
\label{Relationship_gh}
\eeq
where $\hat I$ is the identity matrix and $\hat s$ denotes the matrix representation of $s^\nu{}_{\alpha}$. 

For the purely pseudo-electric VEV configuration, substituting the background field into Eq.~\eqref{Relationship_gh} and evaluating the matrix determinant gives
\beq
h^{\mu\nu} = \fc{1}{\lt(1 - \xi_1 b^2 - \fc{\xi_2 b^2}{2}\rt) \lt(1 - \xi_1 b^2\rt)} \, g^{\mu\alpha} \lt[ \lt(1 - \xi_1 b^2\rt) \delta^{\nu}_{\alpha} + \xi_2 b^{\nu\beta} b_{\alpha\beta} \rt].
\eeq
Consequently, the corresponding covariant auxiliary metric is given by
\bal
h_{\mu\nu} = g_{\mu\alpha} \lt[ \lt(1 - \xi_1 b^2 - \fc{\xi_2 b^2}{2}\rt) \delta^{\alpha}_{\nu} - \xi_2 b^{\alpha\beta} b_{\nu\beta} \rt].
\label{hdd}
\eal

\subsection{Metric Field Equation}

Varying the action \eqref{Main_Action} with respect to the metric yields the gravitational field equation
\beq
R_{\mu\nu} (\Gamma) - \fc{1}{2} g_{\mu\nu} R (\Gamma) + \Lambda g_{\mu\nu} = T^\tx{eff}_{\mu\nu},
\label{Equation_metric}
\eeq
where the effective energy-momentum tensor is defined as $T^\tx{eff}_{\mu\nu} = T^{\tx{KR}}_{\mu\nu} + T^{\tx{M}}_{\mu\nu}$. In vacuum, the matter contribution vanishes, i.e., $T^{\tx{M}}_{\mu\nu} = 0$, while the KR field contributes
\bal
T_{\mu\nu}^{\tx{KR}} & = \fc{1}{2} H_{\mu\alpha\beta} H_{\nu}^{\ \alpha\beta} 
- \fc{1}{12} g_{\mu\nu} H^{\alpha\beta\gamma} H_{\alpha\beta\gamma} 
- 2 V' B_{\mu\alpha} B^{\alpha}_{\ \nu} 
- \fc{1}{2} g_{\mu\nu} V \nn \\
&+ \xi_1 \Big[ \fc{1}{2} g_{\mu\nu} B^{\alpha\beta} B_{\alpha\beta} R(\Gamma) 
+ 2 B^{\alpha}_{\ \nu} B_{\mu\alpha} R(\Gamma) 
- B^{\alpha\beta} B_{\alpha\beta} R_{\mu\nu}(\Gamma) \Big] \nn \\
&+ \xi_2 \Big[ \fc{1}{2} g_{\mu\nu} B^{\alpha\beta} B^{\gamma}_{\ \beta} R_{\alpha\gamma}(\Gamma) 
- B^{\alpha\beta} B_{\mu\beta} R_{\nu\alpha}(\Gamma) 
- B^{\alpha}_{\ \mu} B^{\beta}_{\ \nu} R_{\alpha\beta}(\Gamma) 
- B^{\alpha\beta} B_{\nu\beta} R_{\mu\alpha}(\Gamma) \Big],
\eal
where the prime denotes differentiation with respect to the argument of the corresponding function. 

Because the matter fields couple exclusively to the fundamental metric $g_{\mu\nu}$, diffeomorphism invariance of the matter action implies the on-shell conservation law $\nabla^{(g)}_{\mu}T_{\rm M}^{\mu\nu}=0$. By contrast, the effective KR contribution $T_{\mu\nu}^{\rm KR}$ contains nonminimal couplings to the curvature constructed from the independent connection and is therefore not, in general, separately conserved with respect to the Levi-Civita covariant derivative associated with $g_{\mu\nu}$.

Under the pseudo-electric VEV configuration, the gravitational field equations \eqref{Equation_metric} can be further reformulated as
\bal
 R_{\mu\nu}(\Gamma) & = \fc{1}{2} g_{\mu\nu} R(\Gamma) - \Lambda g_{\mu\nu} 
 - 2 V' b_{\mu\alpha} b^{\alpha}_{\ \nu} 
 - \xi_1 \Big( \fc{1}{2} g_{\mu\nu} b^2 R(\Gamma) 
 - 2 b^{\alpha}_{\ \nu} b_{\mu\alpha} R(\Gamma) 
 - b^2 R_{\mu\nu}(\Gamma) \Big) \nn \\
 &+ \xi_2 \Big( \fc{1}{2} g_{\mu\nu} b^{\alpha\beta} b^{\gamma}_{\ \beta} R_{\alpha\gamma}(\Gamma) 
 - b^{\alpha\beta} b_{\mu\beta} R_{\nu\alpha}(\Gamma) - b^{\alpha}_{\ \mu} b^{\beta}_{\ \nu} R_{\alpha\beta}(\Gamma) - b^{\alpha\beta} b_{\nu\beta} R_{\mu\alpha}(\Gamma) \Big).
 \label{Equation_reformulate}
 \eal

\subsection{KR Field Equation}

The equation of motion (EOM) for the KR field is derived by varying the action \eqref{Main_Action} with respect to $B_{\mu\nu}$, resulting in
\beq
\nabla^{\lt(g\rt)}_{\alpha} H^{\alpha\mu\nu} = -6 \xi_1 B^{\mu\nu} R(\Gamma) - 6 \xi_2 B^{\gamma[\nu} R^{\mu]}{}_{\gamma}(\Gamma) + 6 V' B^{\mu\nu}, 
\label{Equation_KR}
\eeq
where $\nabla^{\lt(g\rt)}_{\alpha}$ denotes the covariant derivative associated with the Levi-Civita connection of the spacetime metric $g_{\mu\nu}$.


\section{Black Hole Solutions} \label{Solution}

To investigate the spacetime geometry under this nonzero VEV background, we adopt the standard static and spherically symmetric metric ansatz
\beq
ds_{(g)}^2=-F(r)dt^2+G(r)dr^2 + r^2 d\Omega_2^2,
\label{Metric_ansatz}
\eeq
where $d\Omega_2^2\equiv d\theta^2+\sin^2\theta\,d\phi^2$ denotes the line element on the unit two-sphere. For this metric ansatz, the constant-norm condition $b^{\mu\nu}b_{\mu\nu}=-b^2$ fixes the remaining radial profile as $\tilde E(r)=|b|\sqrt{F(r)G(r)/2}$.

Using the spacetime metric ansatz~\eqref{Metric_ansatz}, the auxiliary metric $h_{\mu\nu}$ is first determined from Eq.~\eqref{hdd}, and the independent connection is subsequently computed via Eq.~\eqref{Gamma_hdd}. Upon defining the dimensionless Lorentz-violating parameters $\ell_1 \equiv \xi_1 b^2$ and $\ell_2 \equiv \xi_2 b^2$, substituting the resulting curvature quantities into Eqs.~\eqref{Equation_reformulate} and \eqref{Equation_KR} yields the explicit field equations
\bseq \label{Equation}
\bal
\fc{2(1-\ell_1)(2\ell_1+\ell_2)F''}{F} 
- (1-\ell_1)(2\ell_1+\ell_2) \fc{F'}{F} \fc{G'}{G} 
- \fc{(1-\ell_1)(2\ell_1+\ell_2)F'^2}{F^2} \nn \\
+ \fc{2(\ell_2 + 2\ell_1(2-2\ell_1-\ell_2))}{r} \fc{F'}{F} 
- \fc{2(2 - \ell_1(2\ell_1+\ell_2))}{r} \fc{G'}{G} 
- \fc{4(1+\ell_1)(1-\ell_1)G}{r^2} \nn \\
+ 4(b^2 V' + \Lambda)(1-\ell_1)G 
+ \fc{2(1+\ell_1)(2-2\ell_1-\ell_2)}{r^2} &= 0, \label{Equation_16a} \\
\fc{2(1-\ell_1)(2\ell_1+\ell_2)F''}{F} 
- (1-\ell_1)(2\ell_1+\ell_2) \fc{F'}{F} \fc{G'}{G} 
- \fc{(1-\ell_1)(2\ell_1+\ell_2)F'^2}{F^2} \nn \\
+ \fc{2(2 - \ell_1(2\ell_1+\ell_2))}{r} \fc{F'}{F} 
- \fc{2(\ell_2 + 2\ell_1(2-2\ell_1-\ell_2))}{r} \fc{G'}{G} 
- \fc{4(1+\ell_1)(1-\ell_1)G}{r^2} \nn \\
+ 4(b^2 V' + \Lambda)(1-\ell_1)G 
+ \fc{2(1+\ell_1)(2-2\ell_1-\ell_2)}{r^2} &= 0, \label{Equation_16b} \\
\fc{2F''}{F} - \fc{F'G'}{FG} - \fc{F'^2}{F^2} 
+ \fc{2}{r}\lt(\fc{F'}{F} - \fc{G'}{G}\rt) 
+ \fc{8\Lambda G}{2-2\ell_1-\ell_2} &= 0, \label{Equation_16c} \\
- \fc{(1-\ell_1)(2\ell_1+\ell_2)}{2} \fc{F''}{F} 
+ \fc{(1-\ell_1)(2\ell_1+\ell_2)}{4} \lt( \fc{F'^2}{F^2} + \fc{F' G'}{F G} \rt) - b^2 V'(1-\ell_1) G \nn \\
- \fc{\ell_2 + 2\ell_1(2-2\ell_1-\ell_2)}{2r} \lt( \fc{F'}{F} - \fc{G'}{G} \rt) 
+ \fc{\ell_1}{r^2} \bigg[ 2(1-\ell_1)G - (2-2\ell_1-\ell_2) \bigg] &= 0. \label{Equation_16d}
\eal   
\eseq

Subtracting Eq.~\eqref{Equation_16b} from Eq.~\eqref{Equation_16a} yields 
\beq
\fc{F'(r)}{F(r)} + \fc{G'(r)}{G(r)} = 0.
\eeq
This relation integrates to
\beq
G(r)=\fc{c_1}{F(r)}, \label{Relationship_FG}
\eeq
where $c_1$ is an integration constant, which can be fixed by a constant rescaling of the temporal coordinate.

Furthermore, subtracting Eq.~\eqref{Equation_16c} from Eq.~\eqref{Equation_16a} and substituting the relation \eqref{Relationship_FG} into it, one obtains
\bal
F(r) =& \fc{c_1}{\sigma^2} + \fc{c_2}{r} - \fc{c_1 b^2 \lambda}{3 (1 + \ell_1)\sigma^2} r^2  
 - \fc{c_1 \Lambda}{(1 + \ell_1)\sigma^2} r^2 + \fc{2 c_1  \Lambda}{3 (1 + \ell_1)(1 - \ell_1)\sigma^4} r^2,
\label{Solution_F}
\eal
where $\sigma^2\equiv(1-\ell_1-\fc{\ell_2}{2})/(1-\ell_1)$, and $c_2$ is an integration constant associated with the black hole mass.

Subsequently, substituting Eqs.~\eqref{Relationship_FG} and \eqref{Solution_F} into the remaining independent field equations determines the on-shell value of $\lambda$,
\beq
\lambda = -\fc{4\Lambda}{b^2} + \fc{2\Lambda}{b^2(1-\ell_1)} + \fc{4\Lambda}{b^2(2-2\ell_1-\ell_2)}. 
\label{lambda}
\eeq
Notably, in the absence of Lorentz violation ($\ell_1 = \ell_2 = 0$), this constraint identically yields $\lambda = 0$ for any value of $\Lambda$, ensuring that the theory smoothly recovers GR. Furthermore, one can straightforwardly verify that the solutions~\eqref{Relationship_FG}, \eqref{Solution_F}  and \eqref{lambda} satisfy the complete set of EOMs.

Substituting Eq.~\eqref{lambda} into Eq.~\eqref{Solution_F} and fixing the temporal normalization by choosing $c_1=\sigma^2$, one obtains
\beq
F(r)
=
1+\fc{c_2}{r}
-\fc{\Lambda}{3(1-\ell_1)}r^2,
\qquad
G(r)=\fc{\sigma^2}{F(r)}.
\label{normalized_metric_functions}
\eeq

An immediate consequence of the exact solutions obtained above is that the two Lorentz-violating parameters $\ell_1$ and $\ell_2$ are not restricted to mutually exclusive branches. This differs from the purely metric formulation of KR gravity, where the compatibility of the static field equations requires $\ell_1\ell_2=0$ for nontrivial solutions \cite{yu2025revisiting}.

The distinction from the purely metric formulation is particularly
transparent when $\Lambda=0$, for which the general solutions
\eqref{normalized_metric_functions} reduce to
\begin{equation}
F(r)=1+\frac{c_2}{r},
\qquad
G(r)=\frac{\sigma^2}{F(r)},
\qquad
\lambda=0.
\label{Schwarzschild_like_sol}
\end{equation}
Using Eq.~\eqref{hdd}, the corresponding auxiliary metric is
\bal
ds_h^2
=
(1-\ell_1)
\left[
-F(r)dt^2
+\frac{\sigma^2}{F(r)}dr^2
+\sigma^2r^2d\Omega_2^2
\right].
\eal
By introducing the rescaled coordinates
$\bar t=\sqrt{1-\ell_1}\,t$, and
$\bar r=\sqrt{1-\ell_1}\,\sigma r$,
the auxiliary metric then takes the Schwarzschild form
\bal
ds_h^2
=
-f(\bar r)d\bar t^2
+\frac{d\bar r^2}{f(\bar r)}
+\bar r^2d\Omega_2^2,
\qquad
f(\bar r)=1+\frac{c_3}{\bar r},
\eal
where
$c_3=c_2\sqrt{1-\ell_1-\ell_2/2}$.
Since the independent connection is the Levi-Civita connection of
$h_{\mu\nu}$, it follows that $R_{\mu\nu}(\Gamma)=R(\Gamma)=0$. In addition, the pseudo-electric vacuum satisfies $H_{\alpha\mu\nu}=0$, $V=0$, and $V'=\lambda=0$, while $T_{\mu\nu}^{\rm M}=0$. Consequently, every term in the metric and KR field equations \eqref{Equation_metric} and \eqref{Equation_KR} vanishes separately, without imposing any specific constraints on $\ell_1$ and $\ell_2$. The Palatini formulation therefore consistently admits solutions with both $\ell_1\neq0$ and $\ell_2\neq0$.


\section{Black Hole Mass}\label{Wald_Mass}

Having obtained the exact solutions \eqref{normalized_metric_functions}, we now determine the  mass of the black hole. In standard GR, the mass is commonly evaluated using the ADM or Komar construction \cite{Carroll2019}. In the present case, however, the standard ADM construction is obstructed by the modified asymptotic structure, whereas the ordinary Komar charge does not include the additional contributions arising from the nonminimal KR couplings \cite{An2024,li2025dyonic,yu2025revisiting}. We therefore employ the Iyer-Wald formalism \cite{wald1993black,iyer1994properties} to derive the conserved mass from the full Lagrangian.

In accordance with the Iyer-Wald formalism \cite{wald1993black,iyer1994properties}, the action of a diffeomorphism-invariant theory can be written in differential-form notation as
\beq
S(\psi)
=
\int \mathbf L(\psi)
=
\int \mathcal L(\psi)\,\boldsymbol{\epsilon},
\eeq
where $\psi=\{g_{\mu\nu},\Gamma^\lambda{}_{\mu\nu},B_{\mu\nu}\}$ collectively denotes all dynamical fields,
$\mathcal L$ is the Lagrangian scalar appearing in
the action~\eqref{Main_Action}, and
$\mathbf L=\mathcal L\,\boldsymbol{\epsilon}$ is the corresponding
Lagrangian 4-form. Here, $\boldsymbol{\epsilon}$ denotes the volume 4-form associated with the physical metric $g_{\mu\nu}$. For the
coordinates and orientation of Eq.~\eqref{Metric_ansatz},
\beq
\boldsymbol{\epsilon}
=
\sqrt{-g}\,dt\wedge dr\wedge d\theta\wedge d\phi.
\label{Lagrangian_form}
\eeq

In the standard phase space, the cosmological constant $\Lambda$ and the dimensionless Lorentz-violating parameters $\ell_i$ are held fixed. The variation of the Lagrangian 4-form is then given by
\bal
\delta \mathbf{L} = \mathbf{E}_\psi \delta\psi + d\bm{\Theta}(\psi, \delta\psi),
\label{variation_Lagrangian} 
\eal
where $\mathbf{E}_\psi=0$ denotes the EOMs and $\bm{\Theta}$ is the symplectic potential 3-form.
 
For an infinitesimal diffeomorphism generated by a vector field
$\zeta^\mu$, the Noether current 3-form is defined by
\beq
\mathbf{J}_\zeta = \bm{\Theta}(\psi, \mathscr{L}_\zeta \psi) - i_\zeta \mathbf{L},
\eeq
where $\mathscr{L}_\zeta$ denotes the Lie derivative along $\zeta$, and $i_\zeta$ is the interior product. Since $\mathbf{J}_\zeta$ is closed on-shell ($d \mathbf{J}_\zeta = 0$), it can be locally expressed as the exact differential of a Noether charge 2-form $\mathbf{Q}_\zeta$, i.e.,
\beq
\mathbf{J}_\zeta = d\mathbf{Q}_\zeta.
\eeq

To perform explicit component calculations, we express the symplectic potential, Noether current, and Noether charge forms in terms of their dual tensor components as $\boldsymbol{\Theta} = \Theta^\mu\boldsymbol{\epsilon}_\mu$, $\mathbf J_\zeta = J_\zeta^\mu\boldsymbol{\epsilon}_\mu$, and $\mathbf Q_\zeta = \frac{1}{2} Q_\zeta^{\mu\nu}\boldsymbol{\epsilon}_{\mu\nu}$, where the dual basis forms are given by $\boldsymbol{\epsilon}_\mu \equiv i_{\partial_\mu}\boldsymbol{\epsilon}$ and $\boldsymbol{\epsilon}_{\mu\nu} \equiv i_{\partial_\nu}i_{\partial_\mu}\boldsymbol{\epsilon}$.

We next evaluate the symplectic potential defined in Eq.~\eqref{variation_Lagrangian}. Since derivatives of the independent fields enter the Palatini action only through $R_{\mu\nu}(\Gamma)$ and $H_{\mu\nu\rho}$, the variation $\delta g_{\mu\nu}$ makes no contribution to $\bm{\Theta}$. Retaining the total derivatives generated by $\delta\Gamma^{\lambda}{}_{\mu\nu}$ and $\delta B_{\mu\nu}$,  the symplectic potential takes the form
\bal
\Theta^\mu = \Theta_{\Gamma}^\mu + \Theta_\tx{KR}^\mu 
= \fc{1}{2\kappa} \lt( \mc{K}^{\alpha\beta} \delta\Gamma^\mu{}_{\alpha\beta} 
- \mc{K}^{\alpha\mu} \delta\Gamma^\beta{}_{\alpha\beta} \rt) 
- \fc{1}{6\kappa} H^{\mu\alpha\beta} \delta B_{\alpha\beta},
\label{SP}
\eal
where $\mc{K}^{\mu\nu} = \lt(1 - \xi_1 b^2\rt) g^{\mu\nu} + \xi_2 B^{\mu\gamma} B^{\nu}{}_{ \gamma}$.

Specializing to a variation induced by the Killing vector $\zeta^\mu$, and substituting the following Lie derivatives
\bal
\delta_\zeta \Gamma^\lambda{}_{\mu\nu} &= \nabla^{(\Gamma)}_{\mu} \nabla^{(\Gamma)}_{\nu} \zeta^\lambda + R^\lambda{}_{\nu\rho\mu} \zeta^\rho,  \\
\delta_\zeta B_{\mu\nu} &= \zeta^\rho \nabla^{(g)}_{\rho} B_{\mu\nu} + B_{\rho\nu} \nabla^{(g)}_{\mu} \zeta^\rho + B_{\mu\rho} \nabla^{(g)}_{\nu} \zeta^\rho,
\eal
into Eq.~\eqref{SP}, the Noether current is
\bal
J_\zeta^\mu = \Theta^\mu(\delta_\zeta)-\zeta^\mu\mc{L}.
\eal
On shell, it is exact and can be written as
$J_\zeta^\mu=\nabla^{(g)}_{\nu}Q_\zeta^{\mu\nu}$, which yields
\bal
Q^{\mu\nu}_\zeta = \fc{1}{2\kappa} \lt( \mc{K}^{\beta\nu} \nabla^{(\Gamma)}_{\beta} \zeta^\mu - \mc{K}^{\beta\mu} \nabla^{(\Gamma)}_{\beta} \zeta^\nu \rt) + \fc{1}{3\kappa} H^{\mu\nu\beta} B_{\beta\rho} \zeta^\rho.
\label{NC}
\eal

The Iyer-Wald surface-charge density is a spacetime 2-form that depends linearly on the field perturbations $\delta\psi$ and is defined by
\bal
\mathbf k_\zeta
\equiv
\delta\mathbf Q_\zeta-i_\zeta\mathbf\Theta(\psi,\delta\psi).
\label{Surface_charge_density}
\eal
To compute this, we evaluate Eqs.~\eqref{SP} and \eqref{NC} on a two-sphere at fixed $r$ within the static region. For $\Lambda\leq0$, this region is the black-hole exterior, $r_h<r<\infty$, whereas for $\Lambda>0$, it is the static patch bounded by the black-hole horizon $r_h$ and the cosmological horizon $r_c$, with $r_h<r<r_c$. Horizon values are understood as limits taken from within the corresponding static region. The resulting expressions are
\bal
\mathbf{Q}_\zeta &= \fc{2 - 2\ell_1 - \ell_2}{4\kappa } \fc{F'}{\sqrt{F G}} \, r^2 \, \sin\theta \, d\theta \wedge d\phi,     \\
i_\zeta \mathbf{\Theta} &= \fc{2 - 2\ell_1 - \ell_2}{2\kappa} \bigg[ - \fc{F'}{4 F^{3/2} \sqrt{G}} \, \delta F - \fc{\sqrt{F}}{r G^{3/2}} \, \delta G  \nn \\
& - \fc{F'}{4 \sqrt{F} G^{3/2}} \, \delta G + \fc{1}{2 \sqrt{F G}} \, \delta F' \bigg] \, r^2 \, \sin\theta \, d\theta \wedge d\phi,     \\
\mathbf k_\zeta &= -\fc{2 - 2\ell_1 - \ell_2}{2\kappa} \fc{r}{\sqrt{FG}} \, \delta F \, \sin\theta \, d\theta \wedge d\phi. \label{Boundary_integral}
\eal

At fixed $\Lambda$, substituting the variation of Eq.~\eqref{normalized_metric_functions} into Eq.~\eqref{Boundary_integral} gives, on any admissible sphere $S_R$,
\bal
\delta M
&\equiv\oint_{S_R}\mathbf k_\zeta
=
2\pi \fc{2\ell_1 + \ell_2-2}{\kappa \sigma}\delta c_2
=-\frac{\rho}{2}\delta c_2,
\label{Variation_Wald_mass}
\eal
where $\sigma =\sqrt{(1-\ell_1-\fc{\ell_2}{2})/(1-\ell_1)}$ and $\rho \equiv \sqrt{(1-\ell_1) \lt(1-\ell_1-\fc{\ell_2}{2}\rt)}$. The result is independent of $\Lambda$. For $\Lambda\leq0$, it agrees with the asymptotic charge, and for $\Lambda>0$, it defines the Hamiltonian charge in the static patch.

Integrating Eq.~\eqref{Variation_Wald_mass} and fixing the integration constant by requiring $M=0$ for $c_2=0$, the conserved mass is uniquely determined as
\bal
M = -\fc{\rho}{2} c_2.
\label{physical_mass}
\eal

Finally, substituting $c_2=-2M/\rho$ into
Eq.~\eqref{normalized_metric_functions}, the spacetime metric
\eqref{Metric_ansatz} becomes
\bal
ds_{(g)}^2= -\lt(1-\fc{2M}{\rho r}-\fc{\sigma\Lambda}{3\rho}r^2 \rt)dt^2
+\fc{\sigma^2}{1-\fc{2M}{\rho r} -\fc{\sigma\Lambda}{3\rho}r^2}dr^2
+r^2\lt(d\theta^2+\sin^2\theta\,d\phi^2\rt).
\label{Eq_Solution}
\eal
In the Lorentz-invariant limit $\ell_1=\ell_2=0$, it recovers the standard Schwarzschild-(A)dS spacetime.

The dimensionless parameters $\ell_1$ and $\ell_2$ characterize the Lorentz-violating corrections induced by the nonzero VEV of the KR field. The Kretschmann scalar of the obtained spacetime reads
\beq
R^{\alpha\beta\gamma\delta}R_{\alpha\beta\gamma\delta} = \fc{24(1+\sigma^4)M^2}{\rho^2\sigma^4 r^6} + \fc{4(1+\sigma^4)\Lambda^2}{3\rho^2\sigma^2}.
\eeq
The explicit dependence of this curvature invariant on the Lorentz-violating parameters indicates that the deformation induced by the KR background is a genuine modification of the spacetime geometry, rather than a coordinate effect.


\section{Thermodynamics}\label{Thermodynamics}

Having determined the conserved mass via the Iyer-Wald formalism, we now examine the thermodynamic consistency of the black hole solutions. Although the metric admits an asymptotically de Sitter branch ($\Lambda>0$), a generic asymptotically de Sitter black hole possesses both a black-hole horizon and a cosmological horizon, which are generally associated with distinct Hawking temperatures. Although various approaches to de Sitter black hole thermodynamics have been proposed in the literature \cite{Mbarek2019,Haroon2020}, a well-defined global equilibrium formulation for such generic two-horizon spacetimes remains elusive. Consequently, we restrict our analysis to the solutions with $\Lambda\leq0$.

To formulate the extended thermodynamics, we enlarge the solution space by allowing $\Lambda$ to vary while keeping the parameters $\ell_i$ fixed. Although $\Lambda$ is constant within each spacetime, the extended phase space includes neighboring solutions with different values of $\Lambda$ and therefore allows $\delta\Lambda\neq0$. This extension introduces the thermodynamic-volume term without altering the fixed-$\Lambda$ mass obtained above. The Lagrangian variation in the extended phase space is therefore \cite{xiao2024extended}
\beq
\delta\mathbf L
=\mathbf E_\psi\delta\psi
+d\bm{\Theta}(\psi,\delta\psi)
+\frac{\partial\mathbf L}{\partial\Lambda}\delta\Lambda.
\label{extended_variation_Lagrangian}
\eeq

Using the surface-charge density $\mathbf k_\zeta$ introduced in Eq.~\eqref{Surface_charge_density}, the on-shell extended Iyer-Wald identity is 
\beq
d\mathbf k_\zeta=\boldsymbol\omega(\psi;\delta\psi,\mathscr L_\zeta\psi)
+\boldsymbol{\mathcal J}_\Lambda,
\label{local_IW_identity}
\eeq
where $\boldsymbol{\mathcal J}_\Lambda\equiv-i_\zeta
\left(\frac{\partial\mathbf L}{\partial\Lambda}\right)\delta\Lambda$, and $\boldsymbol{\omega}(\psi;\delta_1\psi,\delta_2\psi)\equiv\delta_1\bm{\Theta}(\psi,\delta_2\psi)-\delta_2\bm{\Theta}(\psi,\delta_1\psi)$ is the symplectic current 3-form. 

For the stationary solution, $\zeta=\partial_t$ is an exact symmetry, so that $\mathscr L_\zeta\psi=0$ and the symplectic current $\boldsymbol\omega$ vanishes. Consequently, integrating over the hypersurface $\Sigma_R$ bounded by the horizon and an arbitrary two-sphere $S_R$ in the static exterior gives
\beq
\oint_{S_R}\mathbf k_\zeta-\oint_{r_h}\mathbf k_\zeta
=\int_{\Sigma_R}\boldsymbol{\mathcal J}_\Lambda,
\label{IW_identity}
\eeq

Varying the horizon condition $F(r_h)=0$ gives $\delta F(r_h)+F'(r_h)\delta r_h=0$. Using this relation in Eq.~\eqref{Boundary_integral} and integrating over the horizon two-sphere yields
\beq
\delta\mathcal H_h \equiv \oint_{r_h}\mathbf k_\zeta 
= \fc{\rho}{2}F'(r_h)r_h\delta r_h.
\label{horizon_Hamiltonian}
\eeq
For the metric ansatz \eqref{Metric_ansatz}, the surface gravity $\kappa_h$ associated with $\zeta=\partial_t$ gives the Hawking temperature 
\beq
\mathcal T
=\fc{\kappa_h}{2\pi}
=\frac{1}{4\pi}\lim_{r\to r_h}\frac{F'(r)}{\sqrt{F(r)G(r)}}
=\frac{F'(r_h)}{4\pi\sigma}
=\frac{1}{4\pi\sigma}
\left( \frac{1}{r_h} - \frac{\sigma\Lambda}{\rho}r_h \right),
\eeq
whereas the Wald entropy is
\beq
\mathcal S
=
\frac{1}{\mathcal T}
\oint_{r_h}\mathbf Q_\zeta
=
\pi\rho\sigma r_h^2 .
\eeq
This gives
$\delta\mathcal S=2\pi\rho\sigma r_h\delta r_h$. Consequently,
Eq.~\eqref{horizon_Hamiltonian} gives
\beq
\delta\mathcal H_h=\mathcal T\delta\mathcal S.
\label{horizon_Hamiltonian_2}
\eeq

For the same extended variation, the surface contribution on $S_R$ is
\beq
\delta\mathcal H_R
\equiv\oint_{S_R}\mathbf k_\zeta
=\delta M+\frac{\sigma R^3}{6}\delta\Lambda.
\label{extended_outer_boundary}
\eeq

For the action \eqref{Main_Action}, $\partial\mathbf L/\partial\Lambda=-\boldsymbol\epsilon/\kappa$, and hence
\beq
\boldsymbol{\mathcal J}_{\Lambda}
=
-i_{\zeta}
\left(
\frac{\partial\mathbf L}{\partial\Lambda}
\right)\delta\Lambda
=\frac{\sqrt{-g}}{\kappa}\delta\Lambda\,
dr\wedge d\theta\wedge d\phi.
\eeq
With $\sqrt{-g}=\sigma r^2\sin\theta$, the bulk term is evaluated as
\beq
\int_{\Sigma_R}\boldsymbol{\mathcal J}_\Lambda
=\frac{4\pi\sigma}{3\kappa}(R^3-r_h^3)\delta\Lambda
=\frac{\sigma}{6}(R^3-r_h^3)\delta\Lambda.
\label{extended_bulk_term}
\eeq

Substituting Eqs.~\eqref{horizon_Hamiltonian_2},
\eqref{extended_outer_boundary}, and
\eqref{extended_bulk_term} into Eq.~\eqref{IW_identity}, the terms
proportional to $R^3$ cancel, leaving
\beq
\delta M-\mathcal T\delta\mathcal S
=-\frac{\sigma}{6}r_h^3\delta\Lambda.
\eeq
Identifying the bare cosmological constant with the thermodynamic
pressure, $P=-\Lambda/(8\pi)$, we obtain the first law in the extended phase space
\beq
\delta M = \mathcal T\delta\mathcal S + \mathcal V\delta P,
\label{first_law}
\eeq
where $\mathcal V=\frac{4\pi\sigma}{3}r_h^3$ is the thermodynamic volume conjugate to $P$. Thus, $M$ is identified as the thermodynamic enthalpy.

Furthermore, solving the horizon condition $F(r_h)=0$ gives the mass $M$ in terms of the horizon radius $r_h$,
\beq
M = \frac{\rho}{2}r_h - \frac{\sigma\Lambda}{6}r_h^3 .
\label{M_rh}
\eeq
Direct substitution of the thermodynamic quantities obtained above
then verifies the Smarr relation
\beq
M = 2\left( \mathcal T\mathcal S - \mathcal V P \right),
\eeq
Therefore, despite the modifications arising from the Lorentz-violating KR background, the black hole retains the standard extended thermodynamic structure.


\section{Solar System tests} \label{Tests}

To constrain the Lorentz-violating parameters, we confront the solution with precision Solar System observations. Since the cosmological constant is negligible on these scales, we set $\Lambda=0$ in Eq.~\eqref{Eq_Solution}.

It is important to distinguish the conserved mass $M$ obtained from the Iyer-Wald formalism from the mass parameter inferred from Solar System dynamics. Planetary ephemerides fit the heliocentric gravitational parameter, rather than the solar mass independently of the Newtonian gravitational constant \cite{fienga2011inpop,fienga2024testing}.  For the present metric, its geometrized value follows from the Newtonian potential $\Phi_{\rm N}=-M/(\rho r)$ and is therefore $M/\rho$, rather than $M$ itself. We thus introduce the orbital mass parameter
\beq
\bar{M}\equiv \fc{M}{\rho}, \label{Keplerian_mass}
\eeq
which we refer to as the Keplerian mass parameter.

For $\Lambda=0$, the metric function becomes $F(r)=1-2\bar M/r$ and $G(r)=\sigma^2/F(r)$. The remaining Lorentz-violating deformation is encoded in $\sigma^2=(1-\ell_1-\fc{\ell_2}{2})/(1-\ell_1)=1-q/2$, where $q\equiv\ell_2/(1-\ell_1)$ parametrizes the residual radial deformation after the central mass parameter $\bar M$ has been calibrated. Consequently, the Solar System tests considered below constrain $q$, but cannot determine $\ell_1$ and $\ell_2$ separately.

The trajectory of a test particle along a geodesic is governed by the corresponding Lagrangian
\beq
\chi = -\fc{1}{2} g_{\alpha\beta} \dot{x}^\alpha \dot{x}^\beta,   \label{Lagrangian}
\eeq
where the overdot denotes differentiation with respect to an affine parameter $\omega$. The normalization condition gives $\chi=\eta/2$, with $\eta=1$ for massive particles, for which $\omega$ can be identified with the proper time $\tau$, and $\eta=0$ for massless photons.

Substituting the metric \eqref{Eq_Solution} into the Lagrangian \eqref{Lagrangian} gives
\beq
\chi = \fc{1}{2} \lt( F(r) \dot{t}^2 - \sigma^2 F(r)^{-1} \dot{r}^2 - r^2 \dot{\theta}^2 - r^2 \sin^2\theta \dot{\phi}^2 \rt).
\eeq
Variation with respect to $\theta$ yields the Euler-Lagrange equation
\beq
r^2 \ddot{\theta} + 2r \dot{r} \dot{\theta} - r^2 \dot{\phi}^2 \sin\theta \cos\theta = 0.
\eeq
For the initial conditions $\theta=\fc{\pi}{2}$ and $\dot{\theta}=0$, this equation implies $\ddot{\theta}=0$. Consequently, the particle's motion remains confined to the equatorial plane. Under this condition, the normalization equation reduces to
\beq
F(r) \lt( \fc{dt}{d\omega} \rt)^2 - \sigma^2 F(r)^{-1} \lt( \fc{dr}{d\omega} \rt)^2 - r^2 \lt( \frac{d\phi}{d\omega} \rt)^2 = \eta.  \label{Equation_Lagrangian}
\eeq

The invariance of the Lagrangian under time translations and spatial rotations about the symmetry axis gives rise to two conserved quantities, associated with the Killing vector fields $\partial_t$ and $\partial_\phi$, i.e.,
\beq
E = \fc{\pt \chi}{\pt \dot{t}} = F(r) \fc{dt}{d\omega},  \label{E}
\eeq
\beq
L = -\fc{\pt \chi}{\pt \dot{\phi}} = r^2 \fc{d\phi}{d\omega}.  \label{L}
\eeq
For massive particles with $\omega=\tau$, $E$ and $L$ denote the energy and angular momentum per unit rest mass, respectively. For photons, they are affine-normalized conserved quantities.

Having established the essential geodesic equations, we now turn to investigating the three classical tests of gravity within the Solar System: the perihelion precession of Mercury, the deflection of light by the Sun, and the Shapiro time delay.

\subsection{Mercury's perihelion precession}
With Mercury treated as a test particle, we explicitly identify the affine parameter $\omega$ with the proper time $\tau$. Substituting Eqs.~\eqref{Equation_Lagrangian}, \eqref{E}, and \eqref{L} into the geodesic equation then yields the orbital equation
\beq
\lt[ \fc{d}{d\phi} \lt(\fc{1}{r}\rt) \rt]^2 = \fc{E^2}{\sigma^2 L^2} - \fc{F(r)}{\sigma^2 L^2} \lt( 1 + \fc{L^2}{r^2} \rt).
\eeq
By defining the dimensionless variable $u = \fc{L^2}{\bar{M} r}$ and differentiating with respect to the azimuthal angle $\phi$, we obtain
\beq
\fc{d^2 u}{d\phi^2} + \fc{u}{\sigma^2} - \fc{1}{\sigma^2} = \fc{3\bar{M}^2}{\sigma^2 L^2} u^2.
\label{Mercury_Binet}
\eeq

Since $3\bar M^2/L^2\ll1$ in the weak-field regime, we expand the orbit in this quantity while retaining the dependence on $\sigma$ exactly. Accordingly, we write $u(\phi) \approx u_0(\phi)+u_1(\phi)$, where $u_1$ is of order $\bar M^2/L^2$. Substituting this expansion into Eq.~\eqref{Mercury_Binet} and retaining terms through first order in $\bar M^2/L^2$ gives
\bal
\fc{d^2 u_0}{d\phi^2} + \fc{u_0}{\sigma^2} - \fc{1}{\sigma^2} &= 0,  \label{zeroth-order} \\
\fc{d^2 u_1}{d\phi^2} + \fc{u_1}{\sigma^2} &= \fc{3\bar{M}^2}{\sigma^2 L^2} u_0^2. \label{first-order}
\eal

The solution of the zeroth-order equation \eqref{zeroth-order} is
\beq
u_0 = 1 + e \cos \lt(\fc{\phi}{\sigma} \rt),
\label{Zeroth_Solution_recession}
\eeq
where the integration constants are set by the initial condition
$\phi_0=0$ and the orbital eccentricity $e$. 

By substituting the zeroth-order solution into Eq. \eqref{first-order}, one obtains the first-order solution
\beq
u_1 = \fc{3 \bar{M}^2}{L^2} \lt[ \lt(1 + \fc{e^2}{2}\rt) - \fc{e^2}{6} \cos\lt(\fc{2\phi}{\sigma}\rt) + \fc{e\phi}{\sigma} \sin\lt(\fc{\phi}{\sigma}\rt) \rt].
\label{First_Solution_recession}
\eeq
For the extraction of perihelion advance, the first two terms in the brackets are negligible, as they correspond to a constant radial shift and a bounded periodic correction, neither of which contributes to cumulative orbital effects. In contrast, the third term is secular, growing linearly with $\phi$ and inducing a cumulative phase shift over successive revolutions. Accordingly, only this secular contribution is retained to determine the leading-order perihelion advance.

Retaining the secular part of Eq.~\eqref{First_Solution_recession} and combining it with the zeroth-order solution \eqref{Zeroth_Solution_recession}, we obtain
\beq
u \approx 1 + e \cos\lt(\fc{\phi}{\sigma}\rt) + \fc{3 \bar{M}^2}{\sigma L^2} e\phi \sin\lt(\fc{\phi}{\sigma}\rt).
\eeq
By absorbing this secular term into the phase of the cosine function, the solution can be rewritten in the form of a precessing ellipse
\beq
u \approx 1 + e \cos\lt[ \lt(1 - \fc{3 \bar{M}^2}{L^2}\rt) \fc{\phi}{\sigma} \rt].
\eeq

The orbit is quasi-periodic, with the azimuthal period $\Phi$ required to return to the perihelion given by
\beq
\Phi = \fc{2\pi\sigma}{1 - \fc{3 \bar{M}^2}{L^2}} \approx 2\pi\sigma \lt(1 + \fc{3 \bar{M}^2}{L^2}\rt).
\eeq
Consequently, the coordinate perihelion shift relative to $2\pi$ is
\beq
\Delta \Phi_{\rm KR} \approx \sigma \Delta \Phi_{\rm GR} + 2\pi(\sigma-1),
\label{Mercury_shift_mu}
\eeq
where $\Delta\Phi_{\rm GR}=6\pi\bar M^2/L^2$ denotes the perihelion advance predicted by GR. Furthermore, expanding in $q$ and neglecting the subleading mixed corrections, we obtain
\beq
\Delta \Phi_{\rm KR} \approx \Delta \Phi_{\rm GR} - \fc{\pi q}{2}.
\label{Mercury_shift_q}
\eeq

For Mercury, we employ the GR prediction $\Delta\Phi_{\rm GR}=42.9814^{\prime\prime}/\tx{century}$ and  the observed value $\Delta\Phi_{\rm obs}=(42.9794\pm0.0030)^{\prime\prime}/\tx{century}$ \cite{pitjeva2013relativistic}. Adopting the quoted observational uncertainty, we impose
\beq
\Delta\Phi_{\rm obs}^{\rm Min}
\leq \Delta\Phi_{\rm KR}
\leq \Delta\Phi_{\rm obs}^{\rm Max}.
\label{Mercury_Interval_Condition}
\eeq
With Mercury's orbital period of $87.969$ days, the constraint on the combination $q$ is then derived as
\beq
-7.4 \times 10^{-12} \leq q \leq 3.7 \times 10^{-11}.
\eeq

\subsection{Deflection of light}

Setting $\eta=0$ for massless photons, the exact orbital equation can be derived from Eqs.~\eqref{Equation_Lagrangian}, \eqref{E}, and \eqref{L}
\beq
\lt[ \fc{d}{d\phi} \lt( \fc{1}{r} \rt) \rt]^2 = \fc{E^2}{\sigma^2 L^2} - \fc{F(r)}{\sigma^2 r^2}.
\eeq
Introducing the inverse radial coordinate $u = 1/r$ and differentiating with respect to the azimuthal angle $\phi$ yields the modified Binet equation for light rays
\beq
\fc{d^2 u}{d\phi^2} + \fc{u}{\sigma^2} = \fc{3\bar{M}}{\sigma^2} u^2. \label{Eq_Deflection}
\eeq
In the Lorentz-invariant limit $\sigma=1$, this expression reduces to the standard GR result. 

Since $\bar{M}/r\lesssim\bar{M}/D\ll1$ along a Solar System light ray, the nonlinear term is a first-order relativistic correction. We therefore write $u(\phi)\approx u_0(\phi)+u_1(\phi)$ and omit this term at zeroth order,
obtaining
\beq
\fc{d^2 u_0}{d\phi^2} + \fc{u_0}{\sigma^2} = 0,
\eeq
whose solution is
\beq
u_0 = \fc{1}{D} \sin \lt( \fc{\phi}{\sigma} \rt),
\label{Zeroth_Solution_Deflection}
\eeq
Here $D=L/E$ is the impact parameter defined by the ratio of the conserved angular momentum and energy. 

By substituting the zeroth-order solution into the first-order equation, we obtain
\beq
\fc{d^2 u_1}{d\phi^2} + \fc{u_1}{\sigma^2} = \fc{3 \bar{M}}{D^2 \sigma^2} \sin^2 \lt( \fc{\phi}{\sigma} \rt),
\eeq
which admits the solution
\beq
u_1 = \fc{\bar{M}}{D^2} \lt[ 1 + \cos^2 \lt( \fc{\phi}{\sigma} \rt) \rt].
\label{First_Solution_Deflection}
\eeq

Thus, combining Eqs.~\eqref{Zeroth_Solution_Deflection} and \eqref{First_Solution_Deflection}, the first-order perturbative trajectory of the photon is
\beq
u = \fc{1}{D} \sin \lt( \fc{\phi}{\sigma} \rt) + \fc{\bar{M}}{D^2} \lt[ 1 + \cos^2 \lt( \fc{\phi}{\sigma} \rt) \rt].  
\label{Solution_Deflection}
\eeq

For a light ray incident from and exiting to infinity ($r\to\infty$ or $u\to0$), solving Eq.~\eqref{Solution_Deflection} with $u=0$ within this approximation yields
\beq
\sin \lt( \fc{\phi}{\sigma} \rt) = \fc{ D - \sqrt{D^2 + 8 \bar{M}^2} }{2\bar{M}}.
\eeq
Consequently, to leading order in $\bar{M}/D$, the incident and exit angles are respectively
\bal
\phi_{\tx{in}} &= -\fc{2\sigma\bar{M}}{D},     \\
\phi_{\tx{ex}} &= \fc{2\sigma\bar{M}}{D} + \pi\sigma.
\eal
Therefore, the total deflection of the light reads
\beq
\delta_\tx{KR} = \phi_{\tx{ex}}-\phi_{\tx{in}}-\pi = \sigma \delta_{\rm GR} + (\sigma-1)\pi. 
\label{Background_Subtracted_Deflection}
\eeq
where $\delta_{\rm GR}=4\bar M/D$ is the leading-order GR prediction. Expanding in $q$ and neglecting higher-order corrections, we obtain
\beq
\delta_{\rm KR} \approx \delta_{\rm GR}-\fc{\pi q}{4}.
\label{LV_Deflection}
\eeq

Geodetic very long baseline interferometry (VLBI) measures interstation group delays, from which the parameterized post-Newtonian (PPN) parameter $\gamma$ and an effective solar-deflection amplitude are inferred \cite{titov2015vlbi,lambert2009vlbi}. The result $\gamma=0.99992\pm0.00012$ \cite{lambert2011vlbi} corresponds to the effective observational interval
\beq
\delta_{\rm obs}^{\rm Min,Max}
=\fc{1+\gamma^{\rm Min,Max}}{2}\delta_{\rm GR}.
\eeq
Requiring $\delta_{\rm obs}^{\rm Min}\leq\delta_{\rm KR}\leq\delta_{\rm obs}^{\rm Max}$ and using Eq.~\eqref{LV_Deflection}, with $\delta_{\rm GR}=1.7516687^{\prime\prime}$ for a grazing solar ray, gives the constraint
\beq
-2.2\times10^{-10} \leq q\leq 1.1\times10^{-9}.
\eeq

\subsection{Shapiro time delay}

The Shapiro time delay provides another sensitive Solar System probe of Lorentz violation. Restricting the photon trajectory to the equatorial plane ($\theta=\pi/2$), the null condition $ds_{(g)}^2=0$ gives
\beq
-F(r) dt^2 + \sigma^2 F(r)^{-1} dr^2 + r^2 d\phi^2 = 0.    
\label{Equation_Delay}
\eeq

To evaluate the travel time consistently to first order in $\bar{M}$, it is sufficient to use the zeroth-order light path \eqref{Zeroth_Solution_Deflection}, i.e., $r\sin(\phi/\sigma)=D$, which gives
\beq
r^2 d\phi^2 = \fc{\sigma^2 D^2}{r^2 - D^2} dr^2.
\eeq
Substituting it into Eq.~\eqref{Equation_Delay}, we have
\beq
dt^2 = \fc{\sigma^2}{F(r)} \lt[ \fc{1}{F(r)} + \fc{D^2}{r^2 - D^2} \rt] dr^2.
\eeq
Expanding $F(r)=1-2\bar{M}/r$ to first order in $\bar{M}/r$ gives
\beq
dt \approx \pm \fc{\sigma r\,dr}{\sqrt{r^2-D^2}}\lt(1+\fc{2\bar{M}}{r}-\fc{\bar{M} D^2}{r^3}\rt),
\eeq
where the plus and minus signs correspond to the increasing-$r$ and decreasing-$r$ branches, respectively.

The travel time of a light signal between an emitter at $r_\tx{E}$ and a receiver at $r_\tx{R}$ is given by
\bal
T&\approx \sigma\Bigg[
\sqrt{r_\tx{E}^2-D^2}+\sqrt{r_\tx{R}^2-D^2}+2\bar{M}\ln\fc{
\left(r_\tx{E}+\sqrt{r_\tx{E}^2-D^2}\right)
\left(r_\tx{R}+\sqrt{r_\tx{R}^2-D^2}\right)}{D^2}\nn\\
&-\bar{M}\left(
\fc{\sqrt{r_\tx{E}^2-D^2}}{r_\tx{E}}
+\fc{\sqrt{r_\tx{R}^2-D^2}}{r_\tx{R}}
\right)
\Bigg].
\eal
For a round trip at superior conjunction, let $r_\oplus$ and $r_\tx{R}$ denote the heliocentric radial coordinates of Earth and the receiver, respectively, and let $\bar M_\odot$ denote the solar value of $\bar M$. In the limit $D\ll r_\oplus,r_\tx{R}$, twice the one-way result can be organized as
\bal
T^\tx{tot}
&\simeq\sigma\Bigg\{
2\left(\sqrt{r_\oplus^2-D^2}+\sqrt{r_\tx{R}^2-D^2}\right)
+4\bar M_\odot\left[\ln\fc{4r_\oplus r_\tx{R}}{D^2}-1\right]
\Bigg\}\nn\\
&\equiv\sigma\left(T_0^\tx{tot}+\Delta T_\tx{GR}\right),
\label{Cassini_Light_Time}
\eal
where
$T_0^\tx{tot} \equiv2\left(\sqrt{r_\oplus^2-D^2}+\sqrt{r_\tx{R}^2-D^2}\right)$ is the round-trip propagation time in Minkowski spacetime, and $\Delta T_\tx{GR}\equiv4\bar M_\odot\left[\ln\fc{4r_\oplus r_\tx{R}}{D^2}-1\right]$ represents the Shapiro delay predicted by GR. Subtracting the Minkowski propagation time from the total result gives the KR excess delay
\beq
\Delta T_\tx{KR}
\equiv T^\tx{tot}-T_0^\tx{tot}
=(\sigma-1)T_0^\tx{tot}+\sigma\Delta T_\tx{GR}.
\label{Total_Shapiro_Delay}
\eeq
Expanding to first order in $q$ and using $\Delta T_\tx{GR}\ll T_0^\tx{tot}$, we obtain
\beq
\Delta T_\tx{KR}
\simeq\Delta T_\tx{GR}-\fc{q}{4}T_0^\tx{tot}.
\label{LV_Shapiro_Delay}
\eeq
The Cassini experiment used a coherent two-way radio link, for which a time-dependent propagation delay appears as a fractional shift of the received carrier frequency, $y=d(\Delta T)/dt$ \cite{bertotti2003test}. At leading order, we neglect the slow variations of $r_\oplus$ and $r_\tx{R}$ during conjunction and retain the dominant time dependence through the impact parameter $D(t)$. Differentiating Eq.~\eqref{LV_Shapiro_Delay} therefore gives
\beq
y_\tx{KR} \simeq y_\tx{GR}-\fc{q}{4}
\fc{\partial T_0^\tx{tot}}{\partial D}\dot D
=\left(1-\fc{q}{4}
\fc{\partial_D T_0^\tx{tot}}{\partial_D\Delta T_\tx{GR}}\right)y_\tx{GR},
\eeq
where $y_\tx{GR}=\fc{\partial\Delta T_\tx{GR}}{\partial D}\dot D$ with $\dot D\equiv dD/dt$. The result $\gamma=1+(2.1\pm2.3)\times10^{-5}$ \cite{bertotti2003test} corresponds to the effective observational interval
\beq
y_\tx{obs}^{\rm Min,Max}
=\fc{1+\gamma^{\rm Min,Max}}{2}y_\tx{GR}.
\eeq
Requiring $y_\tx{obs}^{\rm Min}\leq y_\tx{KR} \leq y_\tx{obs}^{\rm Max}$, and using the 2002 conjunction geometry, $r_\oplus\simeq1\,\mathrm{AU}$, $r_\tx{R}\simeq7.43\,\mathrm{AU}$, and $D\simeq1.6R_\odot$ \cite{bertotti2003test}, we obtain the constraint
\beq
-5.5\times10^{-8} \leq q \leq 2.5\times10^{-9}.
\eeq


\section{Conclusions}\label{Conclusions}

In this work, we constructed static and spherically symmetric black holes in Palatini gravity nonminimally coupled to a KR field with a nonzero VEV. Treating the metric and the torsion-free affine connection as independent variables, we derived the modified field equations and obtained an exact family of solutions encompassing the $\Lambda>0$, $\Lambda=0$, and $\Lambda<0$ branches. In contrast to the corresponding purely metric formulation, in which the static vacuum equations restrict the two Lorentz-violating parameters $\ell_1$ and $\ell_2$ to mutually exclusive branches, the Palatini formulation consistently allows both parameters to be nonzero. 

Using the Iyer-Wald formalism, we then determined the conserved mass of the black hole solutions. Restricting the extended phase space analysis to the $\Lambda\leq0$ sector, we verified the first law and the Smarr relation and identified the conserved mass as the thermodynamic enthalpy. Thus, the Lorentz-violating solutions in the Palatini KR gravity retain the standard form of extended black hole thermodynamics.

Finally, we confronted the $\Lambda=0$ solution with Mercury's perihelion precession, solar light deflection, and the Shapiro time delay. Once the central mass parameter $\bar M=M/\rho$ is calibrated by planetary dynamics, the remaining deviations in the exterior geodesic sector depend on the two Lorentz-violating parameters only through $q=\ell_2/(1-\ell_1)$. The resulting direct-comparison bounds are summarized in Table~\ref{Table_Solar_Constraints}. Among these tests, Mercury's perihelion precession gives the most stringent constraint, $-7.4\times10^{-12} \leq q \leq 3.7\times10^{-11}$.

\begin{table}[H]
\centering
\renewcommand{\arraystretch}{1.35}
\begin{tabular}{|c|c|}
\hline
Solar System test & Constraint on $q$ \\
\hline
Mercury's perihelion precession
& $-7.4\times10^{-12}\leq q \leq 3.7\times10^{-11}$ \\
\hline
Solar light deflection
& $-2.2\times10^{-10}\leq q\leq 1.1\times10^{-9}$ \\
\hline
Shapiro time delay
& $-5.5\times10^{-8}\leq q\leq 2.5\times10^{-9}$\\
\hline
\end{tabular}
\caption{Solar System constraints on  $q=\ell_2/(1-\ell_1)$.}
\label{Table_Solar_Constraints}
\end{table}

All three tests constrain the combination $q$ and therefore cannot disentangle $\ell_1$ from $\ell_2$. A coupled perturbation analysis of the metric, connection, and KR field offers a possible route to breaking this degeneracy, because the underlying $\xi_1$ and $\xi_2$ couplings enter the fluctuations through distinct scalar and tensor structures. Such an analysis would also test the dynamical stability of the solution. We leave these questions for future work.


\section*{ACKNOWLEDGMENTS}

This work is supported by the National Natural Science Foundation of China (Grant No.~12475062) and the Natural Science Foundation of Chongqing (Grant No.~CSTB2024NSCQ-MSX0358). Ke Yang also acknowledges the generous hospitality during a visit to the Lanzhou Center for Theoretical Physics, where part of this work was completed.



\begin{thebibliography}{10}

\bibitem{kostelecky1989spontaneous}
V.A.~Kosteleck{\'y} and S.~Samuel, 
\emph{Spontaneous breaking of Lorentz symmetry in string theory}, 
\href{https://doi.org/10.1103/PhysRevD.39.683}
{{Phys. Rev. D} {\bfseries 39} (1989) 683}.

\bibitem{kostelecky1996expectation}
V.~Kostelecky and R.~Potting, 
\emph{Expectation values, Lorentz invariance, and CPT in the open bosonic string}, 
\href{https://doi.org/10.1016/0370-2693(96)00589-8}
{{Phys. Lett. B} {\bfseries 381} (1996) 89} 
[\href{https://arxiv.org/abs/hep-th/9605088}{{\ttfamily arXiv:hep-th/9605088}}].

\bibitem{colladay1998lorentzviolating}
D.~Colladay and V.A.~Kosteleck{\'y}, 
\emph{Lorentz-violating extension of the standard model}, 
\href{https://doi.org/10.1103/PhysRevD.58.116002}
{\emph{Phys. Rev. D} {\bfseries 58} (1998) 116002}.

\bibitem{kostelecky2004gravity}
V.A.~Kosteleck{\'y}, 
\emph{Gravity, Lorentz Violation, and the Standard Model}, 
\href{https://doi.org/10.1103/PhysRevD.69.105009}
{{Phys. Rev. D} {\bfseries 69} (2004) 105009}.

\bibitem{Kostelecky2011}
V.~Kostelecky and N.~Russell, 
\emph{Data tables for Lorentz and CPT violation}, 
\href{https://doi.org/10.1103/RevModPhys.83.11}
{\emph{Rev. Mod. Phys.} {\bfseries 83} (2011) 11} 
[\href{https://arxiv.org/abs/0801.0287}{{\ttfamily arXiv:0801.0287}}].

\bibitem{Tasson2014}
J.D.~Tasson, 
\emph{What do we know about Lorentz invariance?}, 
\href{https://doi.org/10.1088/0034-4885/77/6/062901}
{{Rept. Prog. Phys.} {\bfseries 77} (2014) 062901} 
[\href{https://arxiv.org/abs/1403.7785}{{\ttfamily arXiv:1403.7785}}].

\bibitem{bailey2006signals}
Q.G.~Bailey and A.~Kostelecky, 
\emph{Signals for Lorentz violation in post-Newtonian gravity}, 
\href{https://doi.org/10.1103/PhysRevD.74.045001}
{{Phys. Rev. D} {\bfseries 74} (2006) 045001} 
[\href{https://arxiv.org/abs/gr-qc/0603030}{{\ttfamily arXiv:gr-qc/0603030}}].

\bibitem{bluhm2008constraints}
R.~Bluhm, N.L.~Gagne, R.~Potting and A.~Vrublevskis, 
\emph{Constraints and stability in vector theories with spontaneous Lorentz violation}, 
\href{https://doi.org/10.1103/PhysRevD.77.125007}
{{Phys. Rev. D} {\bfseries 77} (2008) 125007} 
[\href{https://arxiv.org/abs/0802.4071}{{\ttfamily arXiv:0802.4071}}].

\bibitem{altschul2010lorentz}
B.~Altschul, Q.G.~Bailey and A.~Kostelecky, 
\emph{Lorentz violation with an antisymmetric tensor}, \href{https://doi.org/10.1103/PhysRevD.81.065028}
{{Phys. Rev. D} {\bfseries 81} (2010) 065028} 
[\href{https://arxiv.org/abs/0912.4852}{{\ttfamily arXiv:0912.4852}}].

\bibitem{casana2018exact}
R.~Casana, A.~Cavalcante, F.P.~Poulis and E.B.~Santos, 
\emph{Exact Schwarzschild-like solution in a bumblebee gravity model}, 
\href{https://doi.org/10.1103/PhysRevD.97.104001}
{{Phys. Rev. D} {\bfseries 97} (2018) 104001} 
[\href{https://arxiv.org/abs/1711.02273}{{\ttfamily arXiv:1711.02273}}].

\bibitem{maluf2021black}
R.V.~Maluf and J.C.S.~Neves, 
\emph{Black holes with a cosmological constant in bumblebee gravity}, 
\href{https://doi.org/10.1103/PhysRevD.103.044002}
{{Phys. Rev. D} {\bfseries 103} (2021) 044002} 
[\href{https://arxiv.org/abs/2011.12841}{{\ttfamily arXiv:2011.12841}}].

\bibitem{xu2023static}
R.~Xu, D.~Liang and L.~Shao, 
\emph{Static spherical vacuum solutions in the bumblebee gravity model}, 
\href{https://doi.org/10.1103/PhysRevD.107.024011}
{{Phys. Rev. D} {\bfseries 107} (2023) 024011} 
[\href{https://arxiv.org/abs/2209.02209}{{\ttfamily arXiv:2209.02209}}].

\bibitem{Liu2025d}
J.-Z.~Liu, S.-P.~Wu, S.-W.~Wei and Y.-X.~Liu, 
\emph{Exact black hole solutions in bumblebee gravity with lightlike or spacelike VEVS}, 
\href{https://doi.org/10.1007/s11433-026-2961-8}
{{Sci. China Phys. Mech. Astron.} {\bfseries 69} (2026) 270411} 
[\href{https://arxiv.org/abs/2510.16731}{{\ttfamily arXiv:2510.16731}}].

\bibitem{Liu2025e}
J.-Z.~Liu, W.-D.~Guo, S.-W.~Wei and Y.-X.~Liu, 
\emph{Charged spherically symmetric and slowly rotating charged black hole solutions in bumblebee gravity}, 
\href{https://doi.org/10.1140/epjc/s10052-025-13859-x}
{\emph{Eur. Phys. J. C} {\bfseries 85} (2025) 145} 
[\href{https://arxiv.org/abs/2407.08396}{{\ttfamily arXiv:2407.08396}}].

\bibitem{ding2021slowly}
C.~Ding and X.~Chen, 
\emph{Slowly rotating Einstein-bumblebee black hole solution and its greybody factor in a Lorentz violation model},
\href{https://doi.org/10.1088/1674-1137/abce51}
{{Chin. Phys. C} {\bfseries 45} (2021) 025106} 
[\href{https://arxiv.org/abs/2008.10474}{{\ttfamily arXiv:2008.10474}}].

\bibitem{ding2023rotating}
C.~Ding, Y.~Shi, J.~Chen, Y.~Zhou, C.~Liu and Y.~Xiao, 
\emph{Rotating BTZ-like black hole and central charges in Einstein-bumblebee gravity}, 
\href{https://doi.org/10.1140/epjc/s10052-023-11761-y}
{{Eur. Phys. J. C} {\bfseries 83} (2023) 573} 
[\href{https://arxiv.org/abs/2302.01580}{{\ttfamily arXiv:2302.01580}}].

\bibitem{gullu2022schwarzschildlike}
{\.I}.~G{\"u}ll{\"u} and A.~{\"O}vg{\"u}n, 
\emph{Schwarzschild-like black hole with a topological defect in bumblebee gravity}, 
\href{https://doi.org/10.1016/j.aop.2021.168721}
{{Annals of Physics} {\bfseries 436} (2022) 168721} 
[\href{https://arxiv.org/abs/2012.02611}{{\ttfamily arXiv:2012.02611}}].

\bibitem{li2025dyonic}
S.~Li, L.~Liang and L.~Ma, 
\emph{Dyonic RN-like and Taub-NUT-like black holes in Einstein-bumblebee gravity}, 
\href{https://doi.org/10.1088/1475-7516/2026/03/005}
{{JCAP} {\bfseries 03} (2026) 005} 
[\href{https://arxiv.org/abs/2510.04405}{{\ttfamily arXiv:2510.04405}}].

\bibitem{chen2025taubnutlike}
Y.-Q.~Chen and H.-S.~Liu, 
\emph{Taub-NUT-like black holes in Einstein-bumblebee gravity},
\href{https://doi.org/10.1103/wmhj-83x3}
{{Phys. Rev. D} {\bfseries 112} (2025) 084040} 
[\href{https://arxiv.org/abs/2505.23104}{{\ttfamily arXiv:2505.23104}}].

\bibitem{mai2023extendeda}
Z.-F.~Mai, R.~Xu, D.~Liang and L.~Shao, 
\emph{Extended thermodynamics of the bumblebee black holes}, 
\href{https://doi.org/10.1103/PhysRevD.108.024004}
{{Phys. Rev. D} {\bfseries 108} (2023) 024004} 
[\href{https://arxiv.org/abs/2304.08030}{{\ttfamily arXiv:2304.08030}}].

\bibitem{An2024}
Y.-S.~An, 
\emph{Notes on thermodynamics of Schwarzschild-like bumblebee black hole}, 
\href{https://doi.org/10.1016/j.dark.2024.101520}
{{Phys. Dark Univ.} {\bfseries 45} (2024) 101520} 
[\href{https://arxiv.org/abs/2401.15430}{{\ttfamily arXiv:2401.15430}}].

\bibitem{kanzi2019gup}
S.~Kanzi and {\.I}.~Sakall{\i}, 
\emph{GUP modified Hawking radiation in bumblebee gravity}, 
\href{https://doi.org/10.1016/j.nuclphysb.2019.114703}
{{Nucl. Phys. B} {\bfseries 946} (2019) 114703}.
[\href{https://arxiv.org/abs/1905.00477}{{\ttfamily arXiv:1905.00477}}].

\bibitem{wang2021shadow}
H.-M.~Wang and S.-W.~Wei, 
\emph{Shadow cast by Kerr-like black hole in the presence of plasma in Einstein-bumblebee gravity},  
\href{https://doi.org/10.1140/epjp/s13360-022-02785-6}
{{Eur. Phys. J. Plus} {\bfseries 137} (2022) 571} 
[\href{https://arxiv.org/abs/2106.14602}{{\ttfamily arXiv:2106.14602}}].

\bibitem{Kumar2026}
A.~Kumar, S.U.~Islam and S.G.~Ghosh, 
\emph{Probing Lorentz symmetry violation through lensing observables of rotating black holes}, 
\href{https://doi.org/10.1016/j.dark.2026.102307}
{{Phys. Dark Univ.} {\bfseries 52} (2026) 102307} 
[\href{https://arxiv.org/abs/2509.00127}{{\ttfamily arXiv:2509.00127}}].

\bibitem{cai2023accretiona}
Z.~Cai and R.-J.~Yang, 
\emph{Accretion of the Vlasov gas onto a Schwarzschild-like black hole}, 
\href{https://doi.org/10.1016/j.dark.2023.101292}
{{Phys. Dark Univ.} {\bfseries 42} (2023) 101292} 
[\href{https://arxiv.org/abs/2205.04826}{{\ttfamily arXiv:2205.04826}}].

\bibitem{yang2019effectsa}
R.-J.~Yang, H.~Gao, Y.~Zheng and Q.~Wu, 
\emph{Effects of Lorentz breaking on the accretion onto a Schwarzschild-like black hole}, 
\href{https://doi.org/10.1088/0253-6102/71/5/568}
{{Commun. Theor. Phys.} {\bfseries 71} (2019) 568} 
[\href{https://arxiv.org/abs/1809.00605}{{\ttfamily arXiv:1809.00605}}].


\bibitem{oliveira2021quasinormal}
R.~Oliveira, D.M.~Dantas and C.A.S.~Almeida, 
\emph{Quasinormal frequencies for a black hole in a bumblebee gravity}, 
\href{https://doi.org/10.1209/0295-5075/ac130c}
{{EPL} {\bfseries 135} (2021) 10003} 
[\href{https://arxiv.org/abs/2105.07956}{{\ttfamily arXiv:2105.07956}}].


\bibitem{chen2023quasinormal}
C.~Chen, Q.~Pan and J.~Jing, 
\emph{Quasinormal modes of a scalar perturbation around a rotating BTZ-like black hole in Einstein-bumblebee gravity}, 
\href{https://doi.org/10.1016/j.physletb.2023.138186}
{{Phys. Lett. B} {\bfseries 846} (2023) 138186} 
[\href{https://arxiv.org/abs/2302.05861}{{\ttfamily arXiv:2302.05861}}].

\bibitem{sekhmani2025quasinormal}
Y.~Sekhmani, W.~Liu, W.~Deng and K.~Boshkayev, 
\emph{Quasinormal modes of massive scalar perturbations in slow-rotation bumblebee black holes with traceless conformal electrodynamics}, 
\href{https://doi.org/10.1016/j.dark.2026.102358}
{{Phys. Dark Univ.} {\bfseries 52} (2026) 102358} 
[\href{https://arxiv.org/abs/2510.16639}{{\ttfamily arXiv:2510.16639}}].

\bibitem{liu2023qnms}
W.~Liu, X.~Fang, J.~Jing and J.~Wang, 
\emph{QNMs of slowly rotating Einstein–Bumblebee black hole}, 
\href{https://doi.org/10.1140/epjc/s10052-023-11231-5}
{{Eur. Phys. J. C} {\bfseries 83} (2023) 83} 
[\href{https://arxiv.org/abs/2211.03156}{{\ttfamily arXiv:2211.03156}}].

\bibitem{Wang2022}
Z.~Wang, S.~Chen and J.~Jing, 
\emph{Constraint on parameters of a rotating black hole in Einstein-bumblebee theory by quasi-periodic oscillations}, 
\href{https://doi.org/10.1140/epjc/s10052-022-10475-x}
{{Eur. Phys. J. C} {\bfseries 82} (2022) 528} 
[\href{https://arxiv.org/abs/2112.02895}{{\ttfamily arXiv:2112.02895}}].

\bibitem{ovgun2019exacta}
A.~{\"O}vg{\"u}n, K.~Jusufi and {\.I}.~Sakall{\i}, 
\emph{Exact traversable wormhole solution in bumblebee gravity}, 
\href{https://doi.org/10.1103/PhysRevD.99.024042}
{{Phys. Rev. D} {\bfseries 99} (2019) 024042} 
[\href{https://arxiv.org/abs/1804.09911}{{\ttfamily arXiv:1804.09911}}].

\bibitem{liang2022polarizationsa}
D.~Liang, R.~Xu, X.~Lu and L.~Shao, 
\emph{Polarizations of gravitational waves in the bumblebee gravity model}, 
\href{https://doi.org/10.1103/PhysRevD.106.124019}
{{Phys. Rev. D} {\bfseries 106} (2022) 124019} 
[\href{https://arxiv.org/abs/2207.14423}{{\ttfamily arXiv:2207.14423}}].

\bibitem{amarilo2024gravitational}
K.M.~Amarilo, M.B.F.~Filho, A.A.A.~Filho and J.A.A.S.~Reis, 
\emph{Gravitational waves effects in a Lorentz-violating scenario}, 
\href{https://doi.org/10.1016/j.physletb.2024.138785}
{{Phys. Lett. B} {\bfseries 855} (2024) 138785} 
[\href{https://arxiv.org/abs/2307.10937}{{\ttfamily arXiv:2307.10937}}].

\bibitem{maluf2021bumblebee}
R.V.~Maluf and J.C.S.~Neves, 
\emph{Bumblebee field as a source of cosmological anisotropies}, 
\href{https://doi.org/10.1088/1475-7516/2021/10/038}
{{JCAP} {\bfseries 2021} (2021) 038} 
[\href{https://arxiv.org/abs/2105.08659}{{\ttfamily arXiv:2105.08659}}].

\bibitem{Lai2026}
X.-B.~Lai, Y.-Q.~Dong, Y.-Z.~Fan and Y.-X.~Liu, 
\emph{Stability analysis of cosmological perturbations in the bumblebee model: Parameter constraints and gravitational waves}, 
\href{https://doi.org/10.1103/q6fk-3lkj}
{{Phys. Rev. D} {\bfseries 113} (2026) 044003} 
[\href{https://arxiv.org/abs/2509.13958}{{\ttfamily arXiv:2509.13958}}].

\bibitem{lessa2020modifieda}
L.A.~Lessa, J.E.G.~Silva, R.V.~Maluf and C.A.S.~Almeida, 
\emph{Modified black hole solution with a background Kalb-Ramond field}, 
\href{https://doi.org/10.1140/epjc/s10052-020-7902-1}
{{Eur. Phys. J. C} {\bfseries 80} (2020) 335}.
[\href{https://arxiv.org/abs/1911.10296}{{\ttfamily arXiv:1911.10296}}].

\bibitem{yang2023static}
K.~Yang, Y.-Z.~Chen, Z.-Q.~Duan and J.-Y.~Zhao, 
\emph{Static and spherically symmetric black holes in gravity with a background Kalb-Ramond field}, 
\href{https://doi.org/10.1103/PhysRevD.108.124004}
{{Phys. Rev. D} {\bfseries 108} (2023) 124004} 
[\href{https://arxiv.org/abs/2308.06613}{{\ttfamily arXiv:2308.06613}}].

\bibitem{Liu2024a}
W.~Liu, D.~Wu and J.~Wang, 
\emph{Static neutral black holes in Kalb-Ramond gravity}, 
\href{https://doi.org/10.1088/1475-7516/2024/09/017}
{{JCAP} {\bfseries 09} (2024) 017} 
[\href{https://arxiv.org/abs/2406.13461}{{\ttfamily arXiv:2406.13461}}].

\bibitem{yu2025revisiting}
Z.-X.~Yu, H.-D.~Lyu, M.~Huhe and S.~Li, 
\emph{Revisiting black holes and their thermodynamics in Einstein-Kalb-Ramond gravity}, 
\href{https://doi.org/10.1016/j.physletb.2026.140574}
{{Phys. Lett. B} {\bfseries 878} (2026) 140574} 
[\href{https://arxiv.org/abs/2511.19926}{{\ttfamily arXiv:2511.19926}}].

\bibitem{duan2023electrically}
Z.-Q.~Duan, J.-Y.~Zhao and K.~Yang, 
\emph{Electrically charged black holes in gravity with a background Kalb-Ramond field}, 
\href{https://doi.org/10.1140/epjc/s10052-024-13188-5}
{{Eur. Phys. J. C} {\bfseries 84} (2024) 798} 
[\href{https://arxiv.org/abs/2310.13555}{{\ttfamily arXiv:2310.13555}}].

\bibitem{yang2026charged}
J.-H.~Yang, X.-Y.~Guo, J.-Z.~Liu and Y.-X.~Liu,
\emph{Charged black holes with a Lorentz-violating Kalb-Ramond background},
\href{https://doi.org/10.48550/arXiv.2608.02196}
{{\ttfamily arXiv:2608.02196}}.

\bibitem{liu2025shadowa}
W.~Liu, D.~Wu and J.~Wang, 
\emph{Shadow of slowly rotating Kalb-Ramond black holes}, 
\href{https://doi.org/10.1088/1475-7516/2025/05/017}
{\emph{JCAP} {\bfseries 05} (2025) 017} 
[\href{https://arxiv.org/abs/2407.07416}{{\ttfamily arXiv:2407.07416}}].

\bibitem{Belchior2025}
F.M.~Belchior, R.V.~Maluf, A.Y.~Petrov and P.J.~Porf{\'\i}rio, 
\emph{Global monopole in a Ricci-coupled Kalb-Ramond bumblebee gravity}, 
\href{https://doi.org/10.1140/epjc/s10052-025-14390-9}
{\emph{Eur. Phys. J. C} {\bfseries 85} (2025) 658} 
[\href{https://arxiv.org/abs/2502.17267}{{\ttfamily arXiv:2502.17267}}].

\bibitem{atamurotov2022particle}
F.~Atamurotov, D.~Ortiqboev, A.~Abdujabbarov and G.~Mustafa, 
\emph{Particle dynamics and gravitational weak lensing around black hole in the Kalb-Ramond gravity}, 
\href{https://doi.org/10.1140/epjc/s10052-022-10619-z}
{{Eur. Phys. J. C} {\bfseries 82} (2022) 659}.

\bibitem{kumar2020gravitational}
R.~Kumar, S.G.~Ghosh and A.~Wang, 
\emph{Gravitational deflection of light and shadow cast by rotating Kalb-Ramond black holes}, 
\href{https://doi.org/10.1103/PhysRevD.101.104001}
{{Phys. Rev. D} {\bfseries 101} (2020) 104001}.
[\href{https://arxiv.org/abs/2001.00460}{{\ttfamily arXiv:2001.00460}}].

\bibitem{lessa2021traversable}
L.A.~Lessa, R.~Oliveira, J.E.G.~Silva and C.A.S.~Almeida, 
\emph{Traversable wormhole solution with a background Kalb-Ramond field}, 
\href{https://doi.org/10.1016/j.aop.2021.168604}
{{Annals Phys.} {\bfseries 433} (2021) 168604} 
[\href{https://arxiv.org/abs/2010.05298}{{\ttfamily arXiv:2010.05298}}].

\bibitem{maluf2022exact}
R.V.~Maluf and C.R.~Muniz, 
\emph{Exact solution for a traversable wormhole in a curvature-coupled antisymmetric background field}, 
\href{https://doi.org/10.1140/epjc/s10052-022-10409-7}
{{Eur. Phys. J. C} {\bfseries 82} (2022) 445} 
[\href{https://arxiv.org/abs/2110.12202}{{\ttfamily arXiv:2110.12202}}].

\bibitem{maluf2022bianchi}
R.V.~Maluf and J.C.S.~Neves, 
\emph{Bianchi type I cosmology with a Kalb-Ramond background field}, 
\href{https://doi.org/10.1140/epjc/s10052-022-10109-2}
{{Eur. Phys. J. C} {\bfseries 82} (2022) 135} 
[\href{https://arxiv.org/abs/2111.13165}{{\ttfamily arXiv:2111.13165}}].

\bibitem{Hehl1995}
F.W.~Hehl, J.~McCrea, E.W.~Mielke and Y.~Ne'eman, 
\emph{Metric affine gauge theory of gravity: Field equations, Noether identities, world spinors, and breaking of dilation invariance}, 
\href{https://doi.org/10.1016/0370-1573(94)00111-F}
{\emph{Phys. Rept.} {\bfseries 258} (1995) 1} 
[\href{https://arxiv.org/abs/gr-qc/9402012}{{\ttfamily arXiv:gr-qc/9402012}}].

\bibitem{Pagani2015}
C.~Pagani and R.~Percacci, 
\emph{Quantum gravity with torsion and non-metricity}, 
\href{https://doi.org/10.1088/0264-9381/32/19/195019}
{{Class. Quant. Grav.} {\bfseries 32} (2015) 195019} 
[\href{https://arxiv.org/abs/1506.02882}{{\ttfamily arXiv:1506.02882}}].

\bibitem{filho2023vacuum}
A.A.A.~Filho, J.R.~Nascimento, A.Y.~Petrov and P.J.~Porf{\'\i}rio, 
\emph{Vacuum solution within a metric-affine bumblebee gravity},
\href{https://doi.org/10.1103/PhysRevD.108.085010}
{{Phys. Rev. D} {\bfseries 108} (2023) 085010}   
[\href{https://arxiv.org/abs/2211.11821}{{\ttfamily arXiv:2211.11821}}].

\bibitem{filho2024exact}
A.A.A.~Filho, J.R.~Nascimento, A.Y.~Petrov and P.J.~Porf{\'\i}rio, 
\emph{An exact stationary axisymmetric vacuum solution within a metric-affine bumblebee gravity}, 
\href{https://doi.org/10.1088/1475-7516/2024/07/004}
{{JCAP} {\bfseries 07} (2024) 004}  
[\href{https://arxiv.org/abs/2402.13014}{{\ttfamily arXiv:2402.13014}}].

\bibitem{heidari2024scattering}
N.~Heidari, C.F.B.~Macedo, A.A.A.~Filho and H.~Hassanabadi, 
\emph{Scattering effects of bumblebee gravity in metric-affine formalism},  
\href{https://doi.org/10.1140/epjc/s10052-024-13580-1}
{{Eur. Phys. J. C} {\bfseries 84} (2024) 1221}  
[\href{https://arxiv.org/abs/2407.05321}{{\ttfamily arXiv:2407.05321}}].

\bibitem{jha2024constrain}
S.K.~Jha and A.~Rahaman, 
\emph{Constraining the metric-affine bumblebee gravity parameter using black hole shadows and quasiperiodic oscillations of galactic microquasars}, 
\href{https://doi.org/10.1140/epjc/s10052-025-14110-3}
{{Eur. Phys. J. C} {\bfseries 85} (2025) 441} 
[\href{https://arxiv.org/abs/2409.12909}{{\ttfamily arXiv:2409.12909}}].

\bibitem{gao2024gravitational}
X.-J.~Gao, 
\emph{Gravitational lensing and shadow by a Schwarzschild-like black hole in metric-affine bumblebee gravity}, 
\href{https://doi.org/10.1140/epjc/s10052-024-13338-9}
{{Eur. Phys. J. C} {\bfseries 84} (2024) 973} 
[\href{https://arxiv.org/abs/2409.12531}{{\ttfamily arXiv:2409.12531}}].

\bibitem{filho2025how}
A.A.A.~Filho, 
\emph{How does non-metricity affect particle creation and evaporation in bumblebee gravity?}, 
\href{https://doi.org/10.1088/1475-7516/2025/06/026}
{{JCAP} {\bfseries 2025} (2025) 026} 
[\href{https://arxiv.org/abs/2501.00927}{{\ttfamily arXiv:2501.00927}}].

\bibitem{shi2025neutrinoa}
Y.~Shi and A.A.A.~Filho, 
\emph{Neutrino oscillations induced by a new bumblebee black hole},  
\href{https://arxiv.org/abs/2511.16181}{{\ttfamily arXiv:2511.16181}}.

\bibitem{filho2026gravitational}
A.A.~Ara{\'u}jo~Filho, 
\emph{Gravitational waves in metric-affine bumblebee gravity}, 
\href{https://doi.org/10.1140/epjc/s10052-026-15814-w}
{{Eur. Phys. J. C} {\bfseries 86} (2026) 578} 
[\href{https://arxiv.org/abs/2603.08310}{{\ttfamily arXiv:2603.08310}}].

\bibitem{cabral2026casimir}
D.S.~Cabral, A.A.A.~Filho and A.F.~Santos, 
\emph{Casimir phenomena in bumblebee gravity},  
\href{https://arxiv.org/abs/2603.00850}{{\ttfamily arXiv:2603.00850}}.

\bibitem{bluhm2008spontaneous}
R.~Bluhm, S.-H.~Fung and A.~Kostelecky, 
\emph{Spontaneous Lorentz and diffeomorphism violation, massive modes, and gravity}, 
\href{https://doi.org/10.1103/PhysRevD.77.065020}
{{Phys. Rev. D} {\bfseries 77} (2008) 065020} 
[\href{https://arxiv.org/abs/0712.4119}{{\ttfamily arXiv:0712.4119}}].

\bibitem{Carroll2019}
S.M.~Carroll, \emph{Spacetime and geometry: An introduction to general relativity}, Cambridge University Press, 2nd~ed.(2019).

\bibitem{wald1993black}
R.M.~Wald, 
\emph{Black hole entropy is the Noether charge}, 
\href{https://doi.org/10.1103/PhysRevD.48.R3427}
{{Phys. Rev. D} {\bfseries 48} (1993) R3427}.

\bibitem{iyer1994properties}
V.~Iyer and R.M.~Wald, 
\emph{Some properties of the Noether charge and a proposal for dynamical black hole entropy}, 
\href{https://doi.org/10.1103/PhysRevD.50.846}
{{Phys. Rev. D} {\bfseries 50} (1994) 846} 
[\href{https://arxiv.org/abs/gr-qc/9403028}{{\ttfamily arXiv:gr-qc/9403028}}].

\bibitem{Mbarek2019}
S.~Mbarek and R.B.~Mann, 
\emph{Reverse Hawking-Page phase transition in de Sitter black holes}, 
\href{https://doi.org/10.1007/JHEP02(2019)103}
{{JHEP} {\bfseries 02} (2019) 103} 
[\href{https://arxiv.org/abs/1808.03349}{{\ttfamily arXiv:1808.03349}}].

\bibitem{Haroon2020}
S.~Haroon, R.A.~Hennigar, R.B.~Mann and F.~Simovic, 
\emph{Thermodynamics of Gauss-Bonnet-de Sitter black holes}, 
\href{https://doi.org/10.1103/PhysRevD.101.084051}{\emph{Phys. Rev. D} {\bfseries 101} (2020) 084051} [\href{https://arxiv.org/abs/2002.01567}{{\ttfamily arXiv:2002.01567}}].

\bibitem{xiao2024extended}
Y.~Xiao, Y.~Tian and Y.-X.~Liu,
\emph{Extended black hole thermodynamics from extended Iyer-Wald formalism},
\href{https://doi.org/10.1103/PhysRevLett.132.021401}
{{Phys. Rev. Lett.} {\bfseries 132} (2024) 021401}
[\href{https://arxiv.org/abs/2308.12630}{{\ttfamily arXiv:2308.12630}}].

\bibitem{fienga2011inpop}
A.~Fienga, J.~Laskar, P.~Kuchynka, H.~Manche, G.~Desvignes, M.~Gastineau, I.~Cognard and G.~Theureau,
\emph{The INPOP10a planetary ephemeris and its applications in fundamental physics},
\href{https://doi.org/10.1007/s10569-011-9377-8}
{{Celest. Mech. Dyn. Astron.} {\bfseries 111} (2011) 363}
[\href{https://arxiv.org/abs/1108.5546}{{\ttfamily arXiv:1108.5546}}].

\bibitem{fienga2024testing}
A.~Fienga and O.~Minazzoli,
\emph{Testing theories of gravity with planetary ephemerides},
\href{https://doi.org/10.1007/s41114-023-00047-0}
{{Living Rev. Relativ.} {\bfseries 27} (2024) 1}
[\href{https://arxiv.org/abs/2303.01821}{{\ttfamily arXiv:2303.01821}}].

\bibitem{pitjeva2013relativistic}
E.V.~Pitjeva and N.P.~Pitjev,
\emph{Relativistic effects and dark matter in the Solar system from observations of planets and spacecraft},
\href{https://doi.org/10.1093/mnras/stt695}
{{Mon. Not. R. Astron. Soc.} {\bfseries 432} (2013) 3431}
[\href{https://arxiv.org/abs/1306.3043}{{\ttfamily arXiv:1306.3043}}].

\bibitem{titov2015vlbi}
O.~Titov and A.~Girdiuk,
\emph{Applying the theory of general relativity to reducing geodetic VLBI data},
\href{https://doi.org/10.1051/0004-6361/201424690}
{{Astron. Astrophys.} {\bfseries 574} (2015) A128}
[\href{https://arxiv.org/abs/1501.02782}{{\ttfamily arXiv:1501.02782}}].

\bibitem{lambert2009vlbi}
S.B.~Lambert and C.~Le Poncin-Lafitte,
\emph{Determining the relativistic parameter $\gamma$ using very long baseline interferometry},
\href{https://doi.org/10.1051/0004-6361/200911714}
{{Astron. Astrophys.} {\bfseries 499} (2009) 331}
[\href{https://arxiv.org/abs/0903.1615}{{\ttfamily arXiv:0903.1615}}].

\bibitem{lambert2011vlbi}
S.B.~Lambert and C.~Le Poncin-Lafitte,
\emph{Improved determination of $\gamma$ by VLBI},
\href{https://doi.org/10.1051/0004-6361/201016370}
{{Astron. Astrophys.} {\bfseries 529} (2011) A70}.

\bibitem{bertotti2003test}
B.~Bertotti, L.~Iess and P.~Tortora,
\emph{A test of general relativity using radio links with the Cassini spacecraft},
\href{https://doi.org/10.1038/nature01997}
{{Nature} {\bfseries 425} (2003) 374}.

\end{thebibliography}
\end{document}